\documentclass[twocolumn]{aastex6}

\usepackage[shortlabels]{enumitem}
\usepackage{hyperref}
\usepackage{soul}
\usepackage{listings}
\usepackage{color}
\usepackage{graphicx}
\usepackage[caption=false]{subfig}

\newcommand{\marvin}{\texttt{Marvin}}
\newcommand{\brain}{\texttt{Brain}}
\newcommand{\datamodel}{\texttt{DataModel}}
\newcommand{\cube}{\texttt{Cube}}

\newcommand{\rss}{\texttt{RSS}}
\newcommand{\dmaps}{\texttt{Maps}}
\newcommand{\modelcube}{\texttt{ModelCube}}
\newcommand{\spaxel}{\texttt{Spaxel}}
\newcommand{\bin}{\texttt{Bin}}
\newcommand{\rssfiber}{\texttt{RSSFiber}}
\newcommand{\query}{\texttt{Query}}
\newcommand{\results}{\texttt{Results}}
\newcommand{\plate}{\texttt{Plate}}
\newcommand{\image}{\texttt{Image}}
\newcommand{\sdsstree}{\texttt{sdss-tree}}
\newcommand{\sdssaccess}{\texttt{sdss-access}}
\newcommand{\flask}{\texttt{Flask}}
\newcommand{\sciserver}{\texttt{SciServer}}
\newcommand{\browserstack}{\texttt{BrowserStack}}
\newcommand{\jupyter}{\texttt{Jupyter}}
\newcommand{\Ha}{\ensuremath{\rm H\alpha}}
\newcommand{\Hb}{\ensuremath{\rm H\beta}}

\newcommand{\othree}{\textrm{[O\,{\sc iii}]}}
\newcommand{\otwo}{\textrm{[O\,{\sc ii}]}}
\newcommand{\ntwo}{\textrm{[N\,{\sc ii}]}}
\newcommand{\htwo}{\textrm{H\,{\sc ii}}}

\definecolor{codegreen}{rgb}{0,0.6,0}
\definecolor{codegray}{rgb}{0.5,0.5,0.5}
\definecolor{codepurple}{rgb}{0.58,0,0.82}
\definecolor{backcolour}{rgb}{0.95,0.95,0.92}

\lstdefinestyle{mystyle}{
    backgroundcolor=\color{backcolour},
    commentstyle=\color{codegreen},
    keywordstyle=\color{magenta},
    numberstyle=\tiny\color{codegray},
    stringstyle=\color{codepurple},
    basicstyle=\ttfamily\small,
    breakatwhitespace=false,
    breaklines=false,
    captionpos=b,
    keepspaces=true,
    numbers=none,
    numbersep=5pt,
    showspaces=false,
    showstringspaces=false,
    showtabs=false,
    tabsize=4
}

\begin{document}


\title{Marvin: A Toolkit for Streamlined Access and Visualization of the SDSS-IV MaNGA Data Set}


\author{Brian Cherinka\altaffilmark{1}, Brett H.~Andrews\altaffilmark{2}, Jos\'e S\'anchez-Gallego\altaffilmark{3}, Joel Brownstein\altaffilmark{4}, Mar\'ia Argudo-Fern\'andez\altaffilmark{5,6}, Michael Blanton\altaffilmark{7}, Kevin Bundy\altaffilmark{8}, Amy Jones\altaffilmark{13}, Karen Masters\altaffilmark{10,11}, David R. Law\altaffilmark{1}, Kate Rowlands\altaffilmark{9}, Anne-Marie Weijmans\altaffilmark{12}, Kyle Westfall\altaffilmark{8},  Renbin Yan\altaffilmark{14}}
\altaffiltext{1}{Space Telescope Science Institute, 3700 San Martin Drive, Baltimore, MD 21218, USA}
\altaffiltext{2}{Department of Physics and Astronomy and PITT PACC, University of Pittsburgh, 3941 O’Hara Street, Pittsburgh, PA 15260, USA}
\altaffiltext{3}{Department of Astronomy, Box 351580, University of Washington, Seattle, WA 98195, USA}
\altaffiltext{4}{Department of Physics and Astronomy, University of Utah, 115 S 1400 E, Salt Lake City, UT 84112, USA}
\altaffiltext{5}{Centro de Astronom\'ia (CITEVA), Universidad de Antofagasta, Avenida Angamos 601 Antofagasta, Chile}
\altaffiltext{6}{Chinese Academy of Sciences South America Center for Astronomy, China-Chile Joint Center for Astronomy, Camino El Observatorio, 1515, Las Condes, Santiago, Chile}
\altaffiltext{7}{Department of Physics, New York University, 726 Broadway, New York, NY 10003, USA}
\altaffiltext{8}{University of California Observatories, University of California Santa Cruz, 1156 High Street, Santa Cruz, CA 95064, USA}
\altaffiltext{9}{Department of Physics and Astronomy, Johns Hopkins University, 3400 N. Charles St., Baltimore, MD 21218, USA}
\altaffiltext{10}{Department of Physics and Astronomy, Haverford College, 370 Lancaster Avenue, Haverford, Pennsylvania 19041, USA}
\altaffiltext{11}{Institute of Cosmology \& Gravitation, University of Portsmouth, Dennis Sciama Building, Portsmouth, PO1 3FX, UK}
\altaffiltext{12}{School of Physics and Astronomy, University of St Andrews, North Haugh, St Andrews, KY16 9SS, UK}
\altaffiltext{13}{Department of Physics and Astronomy, University of Alabama, Tuscaloosa, AL 35487, USA}
\altaffiltext{14}{Department of Physics and Astronomy, University of Kentucky, 505 Rose St., Lexington, KY 40506-0057, USA}

\email{bcherinka@stsci.edu}

%



\begin{abstract}

The Mapping Nearby Galaxies at Apache Point Observatory (MaNGA) survey, one of three core programs of the fourth-generation Sloan Digital Sky Survey (SDSS-IV), is producing a massive, high-dimensional integral field spectroscopic data set.  However, leveraging the MaNGA data set to address key questions about galaxy formation presents serious data-related challenges due to the combination of its spatially inter-connected measurements and sheer volume.  For each galaxy, the MaNGA pipelines produce relatively large data files to preserve the spatial correlations of the spectra and measurements, but this comes at the expense of storing the data set in a coarsely-chunked manner.  The coarse chunking and total volume of the data make it time-consuming to download and curate locally-stored data.  Thus, accessing, querying, visually exploring, and performing statistical analyses across the whole data set at a fine-grained scale is extremely challenging using just FITS files.  To overcome these challenges, we have developed \marvin: a toolkit consisting of a Python package, Application Programming Interface (API), and web application utilizing a remote database.  \marvin's robust and sustainable design minimizes maintenance, while facilitating user-contributed extensions such as high level analysis code.  Finally, we are in the process of abstracting out \marvin's core functionality into a separate product so that it can serve as a foundation for others to develop \marvin-like systems for new science applications.

\end{abstract}


\section{Introduction}
\label{sec:intro}

Large astronomy collaborations with dedicated facilities pursuing multi-year surveys are producing massive data sets at furious rates.
The data sets from the current generation of surveys, such as the Sloan Digital Sky Survey (hereafter SDSS; \citealt{york2000, strauss2002}), require more disk space than is available on personal computers and some moderate-sized institution-level servers.  However, the next generation of surveys, such as the Large Synoptic Sky Survey \citep{ivezic2008} and the Square Kilometer Array \citep{braun2015}, will create data sets that will be far too large for all but a few dedicated national-level facilities.
The real power of these immense data sets comes from simultaneously leveraging multiple sources of information (e.g., at different wavelengths) about each object, so connecting the relevant data sources for a comprehensive analysis is critical.
Since individual users cannot store the data locally and need to access portions of the data remotely, bandwidth is often the primary bottleneck.  Speed increases in Internet bandwidth have lagged behind those in computer processors (i.e., Moore's law; \citealt{moore1965}) by 10\% \citep{nielsen1998}; the effect of this lag has compounded over decades, up to the present, to exacerbate the gap. Consequently, only a subset of the data can be transferred. However, selecting this subset often requires access to the whole data set, which requires remote operations, especially queries.

SDSS was one of the earliest and remains one of the strongest driving forces in astronomy pushing the philosophy of public data releases that make astronomy a leader in open science.  Crucially, these data releases are served with robust data distribution systems and come thoroughly documented.  These two often-overlooked aspects have lowered the entry barrier and enabled thousands of professional astronomers and many times more public users to take advantage of this powerful data set.  \marvin\ extends this mission by providing code to facilitate data use by professional astronomers, scientists in other fields (e.g., physics, computer science, and statistics), data scientists, citizen scientists, educators, and students.

The current phase (2014--2020) of SDSS, SDSS-IV \citep{blanton2017}, consists of three simultaneous surveys, including the Mapping Nearby Galaxies at Apache Point Observatory (MaNGA; \citealt{bundy2015}) survey.  Legacy SDSS \citep{york2000} took spectra of only the central regions of galaxies \citep{strauss2002}, whereas MaNGA takes hundreds of spectra per galaxy arranged in a hexagonal grid across the face of the galaxy \citep{drory2015}, using the SDSS/BOSS spectrographs \citep{smee2013} on the SDSS telescope \citep{gunn2006}.  Typically, there are 3 dithered sets of 3 individual exposures offset from each other which are combined into a data cube \citep{law2016, yan2016a, yan2016b}.  Thus, each object is not represented by just a single central spectrum, but rather a well-sampled grid of spectra.

\begin{figure*}
\centering
\includegraphics[width=\textwidth]{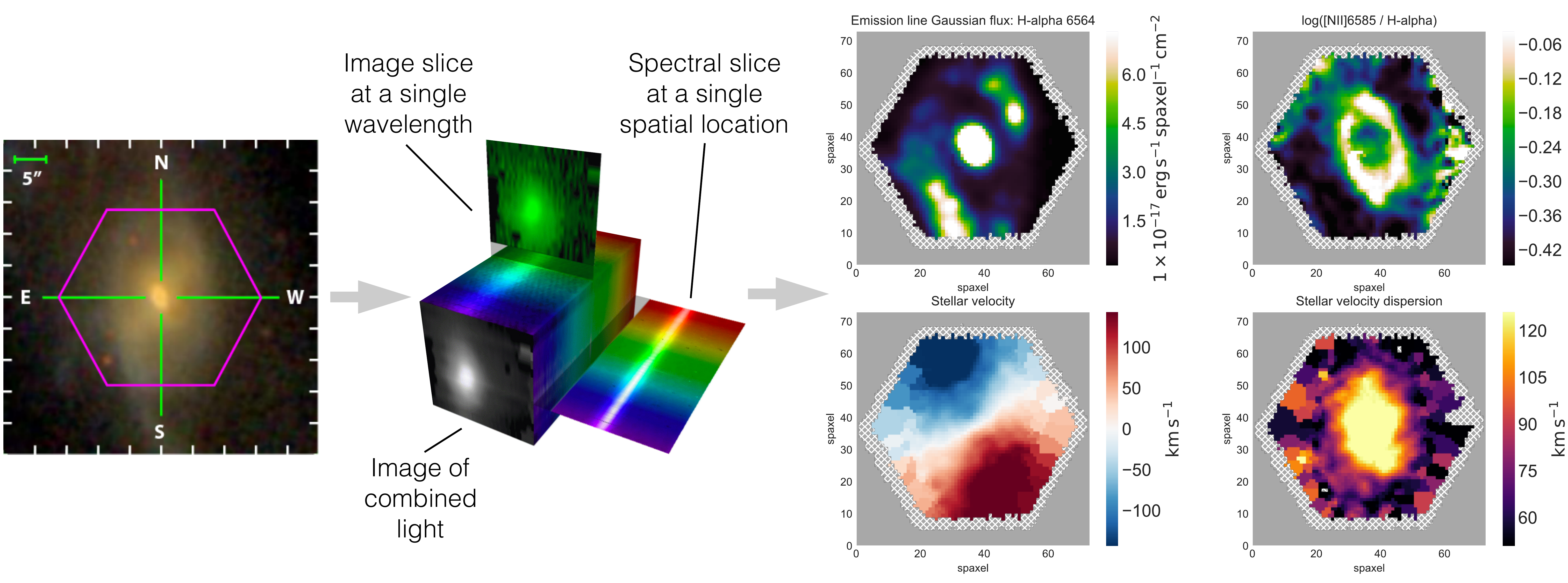}
\caption{\textit{Left}: \textit{gri} image of MaNGA 1-596678 with the IFU field of view shown in purple.  \textit{Middle}: IFU observations produce three-dimensional datacubes, with two spatial and one spectral dimensions (credit: Stephen Todd and Douglas Pierce-Price; IFS Wiki; \url{http://ifs.wikidot.com}).  \textit{Right}: Spectral analysis of individual spaxels produces hundreds of two-dimensional maps for each galaxy spanning a wide range of physical properties.  The four example maps shown for 1-596678 are \Ha\ flux (top left), log(\ntwo\ $\lambda 6585$ / \Ha) flux (top right), stellar velocity (bottom left), and stellar velocity dispersion corrected for instrumental broadening (bottom right).  The map plots were made using the \marvin\ code provided in the \ref{app:map-plotting}.}
\label{fig:datacube}
\end{figure*}

Figure~\ref{fig:datacube} illustrates the format of the MaNGA dataset.   Each data cube consists of two spatial dimensions and one wavelength dimension.  The one-dimensional spectrum at each spatial location can be interpreted in terms of measurements and physical parameters, yielding over 150 two-dimensional maps for each galaxy (Westfall et al.~in prep.), including: gas emission lines, stellar absorption features, stellar surface density, star formation rate surface density, stellar velocities, and gas velocities.  These maps can then be interpreted in terms of global properties of each galaxy: its mass in stars, its mass in dark matter, its total star formation rate, and other quantities.  \marvin\ and the MaNGA maps for 4824 galaxies will be publicly released as part of Data Release 15 (Aguado et al.~2018).

In addition to its complexity, MaNGA's data volume is significant.  MaNGA will observe over 10,000 galaxies \citep{law2015, wake2017}, more than an order of magnitude larger than previous IFU surveys, such as the Atlas$^\mathrm{3D}$ \citep{cappellari2011}, DiskMass \citep{bershady2010}, and CALIFA (Calar Alto Large Integral Field Area; \citealt{sanchez2012}) surveys.  All told, the final MaNGA data release will be 10 terabytes or about 1 gigabyte per galaxy in final summary data products, containing: data cubes and row-stacked spectra in log and linear wavelength sampling, derived analysis maps, and model template data cubes.  Individual data releases contain multiple analyses of each galaxy, each optimized for different science goals, resulting in multiple versions, e.g different binning schemes, of the data cube and maps.  The total volume for all of the MaNGA public data releases will be 35 terabytes due to re-analyses of the same galaxies as the data pipelines improve. Because of these re-analyses, if a given scientific paper is to be replicable, easy access to previous data releases must also be provided.

Further complicating analysis of MaNGA data is its coarsely-chunked storage across separate files for the spectra and derived property maps for each galaxy.  Traditionally data are stored this way to optimize for an object-by-object catalog of files.  This coarsely-chunked data makes querying on MaNGA's spatially-resolved data quite difficult without extensive manual preparation of all files and tracking of correct cross-matches, so queries can only easily be done on global properties.  Exploratory analysis and visualization are cumbersome with coarsely-chunked data, which is compounded by the disconnected packaging of the spectra and maps.  Finally, coarsely-chunked data unnecessarily strains bandwidth and disk space resources because superfluous data need to be transferred and stored.  These challenges encourage traditional object-by-object analyses instead of innovative ones that leverage the statistically significant sample size of MaNGA.

This paper presents software to address these challenges.  Section~\ref{sec:coredesign} describes the initial prototype and its inherent limitations, the core design philosophy of \marvin\ and the components involved.  Section~\ref{sec:marvin-tools} describes the variety of client-based programmatic tools available in \marvin.  Section~\ref{sec:marvin-web} describes the front-facing web portion of \marvin\, which serves as the exploratory portal of entry for new users.  The server-side features and back-end capabilities are discussed in Section~\ref{sec:marvin-backend}.  Section~\ref{sec:usecase} describes a typical science use case for MaNGA and how \marvin\ streamlines its implementation.  In Section~\ref{sec:sustain}, we discuss our current implementation strategy for engaging the long-term sustainability of \marvin.  We summarize and discuss the future potential of \marvin\ in Section~\ref{sec:summary}.   Finally, a series of example code and tutorials are provided in the Appendix.

\section{Core Design}
\label{sec:coredesign}

\subsection{The \marvin\ Prototype}
\label{sec:prototype}
To address the challenge of visually exploring MaNGA data, we developed a prototype version of \marvin\ that existed as a pure web-application.  The prototype displayed optical images, spectra, and property maps for individual galaxies.  These visual displays, in conjunction with a basic annotation system, proved useful for quality assessment of an early version of the MaNGA pipelines.  The prototype also featured a simple query system and provided links to download the FITS data files.

The design choices for the prototype enabled rapid development, but ultimately limited its utility and sustainability.  The images, spectra, and maps were static PNG files, which could not provide the interactive experience required for a complete visual exploration of the complex suite of available parameters.  Queries could only be performed on global properties not local (spatially-resolved) ones.  Data could only be accessed via large files that contained all of the spectra or property maps for a galaxy, making it impossible to retrieve just the spectrum or a single property of an individual spaxel.  Because expanding the feature set of the prototype required creating new static files, the prototype was difficult to extend and time-consuming to maintain.

Furthermore, none of the components in the prototype web-application were usable in a command line form.  Users were forced to reinvent the same visual and search tools if they wanted to use them programmatically.  Such tools could serve as the basis for and be related to advanced programmatic analysis tools.  Every user would end up developing similar tools but within different frameworks, such that each individual's analysis code would not be interoperable with that of other users.

These limitations of the prototype design failed to address any of the inherent challenges of the MaNGA data set.  Thus, a complete redesign and refactor was required to fix these shortcomings, which led to a new design philosophy of \marvin.

\subsection{Design Philosophy and Core Components}
\label{sec:philosophy}
\marvin's design philosophy focuses on eliminating the overhead costs and limitations of accessing the large, coarsely-chunked, and incompletely-linked MaNGA data set.  Solving these issues enables on-demand data access, interactive visual exploration, minimal downloads, spatially-resolved queries, and statistical analyses at a spaxel-level.  \marvin\ provides a feature-rich framework that serves as the building blocks for user-developed analysis tools that can be contributed back into \marvin\ to maximize code reuse and accelerate scientific progress.

\marvin\ is a complete toolkit designed for overcoming the challenges of searching, accessing, and visualizing the MaNGA data.   The core design is centered around a few main components:

\begin{itemize}
	\item A Multi-Modal Access (MMA) system that handles all data flow paths.
	\item An Application Programming Interface (API) based on the Representational State Transfer (REST) architectural style that handles all communication between the client and server.
	\item A \brain, a common core package that handles generic functionalities and abstracts common methods needed during data gathering.
	\item A programmatic \datamodel, that simplifies handling of a large suite of parameters that may differ between data releases and formats.
\end{itemize}

\marvin\ combines and builds on top of these core pieces to provide the following additional tools:

\begin{itemize}
	\item A suite of interconnected Python tools, all based off a core Python tool with the MMA system built-in, with two main tool types:
    	\begin{itemize}
        	\item Data Product Tool: wraps your data products and retrieves specific chunks of data. (e.g., \cube\ or \dmaps\ in \S\ref{sec:galtools})
        	\item Query Tool: performs SQL queries against the remote data, with a pseudo-natural language syntax parser to simplify the user input.
    	\end{itemize}
    	\item A Python \texttt{Interaction} class providing a uniform interface to the API, integrated into all the Tools.
   	\item A web application, built on top of the Tools, for quick data visualization and exploration.
\end{itemize}

These tools work with each other, allowing for multiple entry points into the data, making it easy for users of various domain expertise (i.e from students to power-users), to access the data using the same suite of tools.

\subsection{Multi-Modal Access}
\label{sec:mma}

In the case of MaNGA, the amount of data produced (the final data release will be of order 10~TB) sits on the boundary of what a user can store and analyze locally with normal computing resources.  Future surveys (e.g., the Large Synoptic Survey Telescope) will produce data sets many orders of magnitude larger than MaNGA's, thus requiring the development of new ways to access data.

One of \marvin's core design choices is that data access should be abstracted in a way that makes the origin of the data irrelevant to the final user.  \marvin\ accomplishes this goal with a Multi-Modal Access system with a decision tree that defines what access mode to use and the code implementation that executes it.  Below we describe the data access modes: opening local files, searching local databases, or making API calls to a remote web server.  Each of these data formats carries a series of advantages and disadvantages, but \marvin's MMA allows users to leverage the advantages while minimizing the disadvantages.

Files (e.g. FITS) provide portable data that can be heavily compressed, and they are the current standard for astronomical data distribution.  However, data access can be slow (especially from compressed files), and the data are usually stored in a way that requires a degree of familiarity with the data model.  Moreover, doing searches and cross-analyses between multiple targets usually demands accessing a large number of files and keeping a significant amount of data in memory.

Relational databases solve some of these problems by storing the whole data set in an optimized and well-indexed way, which enables running complex queries efficiently, and provides quicker data access in most situations.  In this case, the main disadvantages are the large size of a monolithic database (comparable to downloading all of the uncompressed files that compose the data set) and the difficulty of learning how to access data, especially compared to access via files.

Finally, data can be stored in servers (either as files or in databases) and accessed remotely via an API call that returns only the subset of data requested in the call.  APIs are convenient for the user since they obviate the need to download data files to a local computer and can be used to abstract the data model. Their main downsides are that the internet is required to access the data and that applications that require access to large amounts of data can be slow to run.

\marvin\ Tools (see Section~\ref{sec:marvin-tools}) include implementations that allow loading data from files, from a database, or via a series of API calls.  However, once the data has been loaded, the Tools behave the same and produce the same results regardless of the data origin.

\begin{figure*}
\centering
\includegraphics[width=\textwidth]{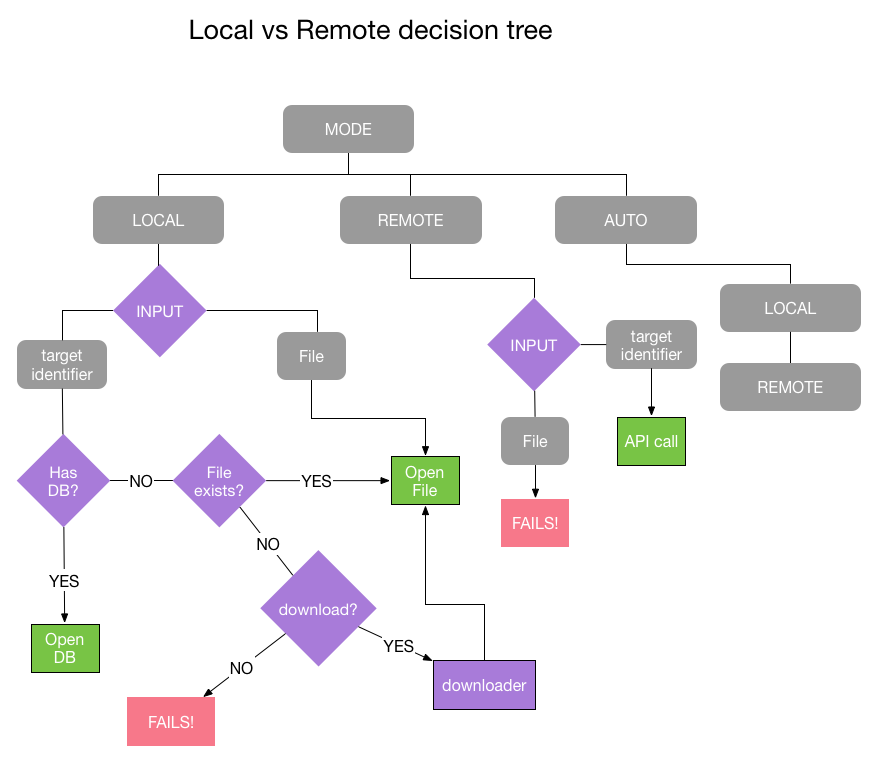}
\caption{The decision tree for the Multi-Modal Access System.  The Multi-Modal Access system operates in three possible modes: local, remote, and auto. See text for a detailed explanation.}
\label{fig:DecTree}
\end{figure*}

Figure~\ref{fig:DecTree} shows the decision tree followed by each tool to decide from where to load data.  If the MMA is being run in ``local'' mode and a target identifier is provided (a plate-IFU or mangaid, which define a unique observation or a single target, respectively; see \citealt{yan2016b}), the code checks if a database is available and, if so, loads the data using it.  If a database cannot be found, the default path file corresponding to that identifier and data release (generated as described in Section~\ref{sec:path-generation}) is used, if the file exists locally.  Alternatively, a file path can be passed to the MMA, in which case that file will be used.

In ``remote'' mode, an API call is done to a remote server with the target identifier and the data release as inputs; the remote server uses the same MMA in ``local'' mode to access the necessary data from a database containing the complete MaNGA data set and returns them.

The default mode for \marvin\ is ``auto'' mode, which tries to access the data in ``local'' mode first and will try in ``remote'' mode upon failure.  This order prioritizes local over remote data access because the former is usually faster, while seamlessly transitioning to the latter if the data is not available locally.  See Appendix~\ref{app:mma} for an illustration of accessing an object with the MMA under different inputs and data origins.

In principle, it would also be possible to set up a system with a complete MaNGA database and use \marvin\ to access it locally.  While setting up such a system would be non-trivial from a technical standpoint, there are situations in which it could be advantageous (e.g., in the case of an institution that wants to provide a local mirror of the MaNGA data set).

\begin{figure*}
\centering
\includegraphics[width=\textwidth]{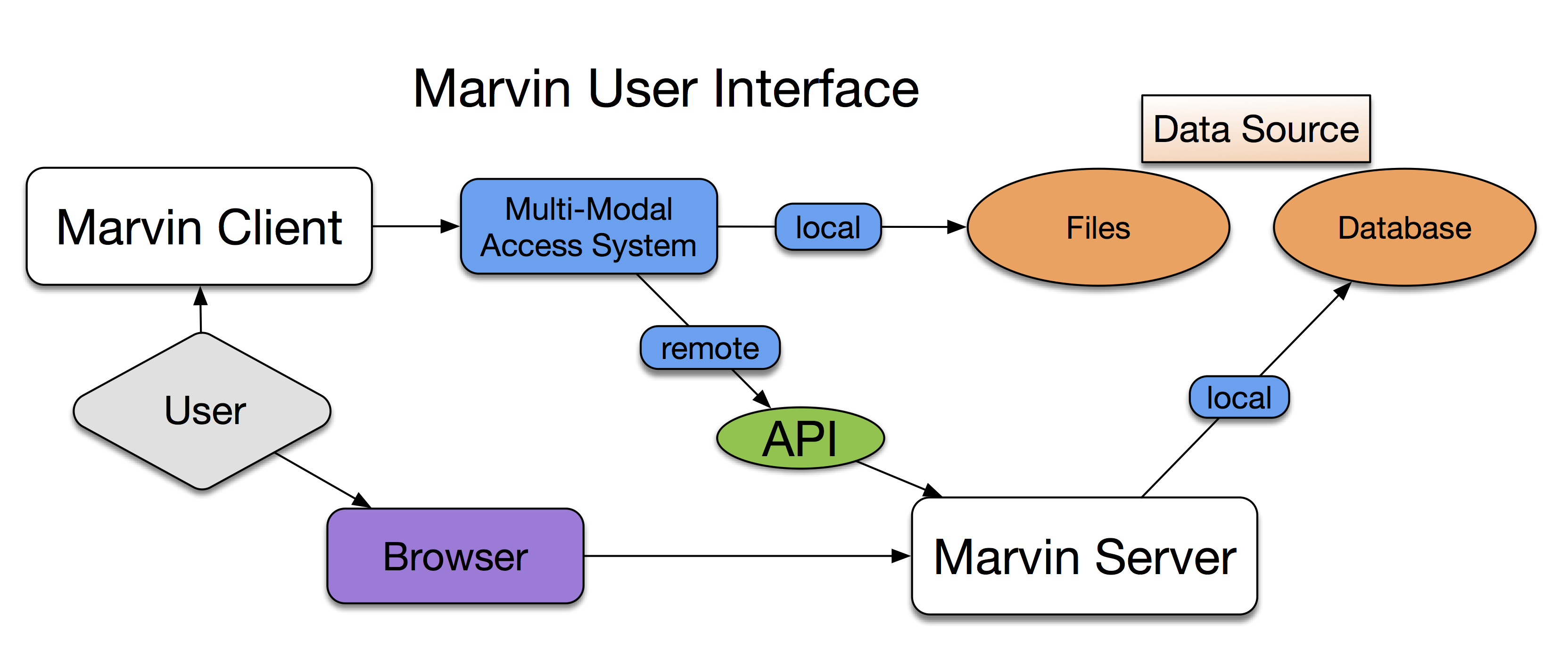}
\caption{A high level user interface of \marvin\ and the Multi-Model Access system, depicting the two major paths of user flow through the system.  Namely, via the browser which communicates directly to the \marvin\ server or via the \marvin\ client, which uses the MMA to decide on local or remote access.}
\label{fig:UserFlow}
\end{figure*}

Figure~\ref{fig:UserFlow} shows a high level overview of the user interface in \marvin.  The user has two main access points: the local \marvin\ client or the web browser interface.  While the browser interface communicates directly with the \marvin\ server, the MMA operating on the client-side decides whether to access data locally or remotely via API calls to the \marvin\ server.  The \marvin\ server (following the MMA decision tree) first attempts to access data from a local database and will fall back to files when needed.

\subsubsection{Abstract Path Generation}
\label{sec:path-generation}

A machine-aware approach to file locations requires generalizing the ability to generate full paths to these files and removing all traces of the base filesystem root directory.  In this way, \marvin\ can be agnostic to whether it is installed on a user's laptop or an SDSS host server.  This layer of functionality is provided by the publicly available \sdssaccess\ \citep{sdssaccess2018} and \sdsstree\ \citep{tree2018} software packages.  \sdsstree\ provides the local system environment variable setup, allowing tools to understand the relative locations of data, while \sdssaccess\ provides a convenient way of navigating local and remote file paths.   Paths to files are defined in a template format, specified with a shortcut name, plus a series of keyword arguments that specify variables within the filenames.  This enables users to specify a robust path to any file simply by adjusting the input variable parameters.  These packages are designed around relative path definitions, allowing a user to replicate a full environment by changing the definition of the base path.  With a single root environment variable set by the user, these packages automatically create a local filesystem structure that mimics the filesystem of the SDSS Science Archive Server hosted at the University of Utah on which the full MaNGA data archive is stored.

For a given file, \sdssaccess\ has the ability to look up the full system path, generate the corresponding HTTP URL, and generate a remote access path for use with \texttt{rsync}.  This flexibility allows \marvin\ to know precisely where to look for a given file locally and also quickly switch to a remote host when needed.  \sdssaccess\ has the ability to download files from an SDSS server using multi-stream \texttt{rsync}, a technology derived from the SDSS Transfer Product \citep{weaver2015}.  This enables fast and robust file transfers, which are particularly helpful for speeding up downloads of many files.  The hierarchy of files is created identically at the destination.  As paths are added to the service, \sdssaccess\ eliminates redundant downloading by first checking for the existence of the file locally and only downloads files that do not currently exist.

\subsection{\marvin's \brain}
\label{sec:brain}

\marvin's \brain\ is a core product that \marvin\ relies on and contains the management and overhead needed for regular tasks.  There are many skills, tasks, and functionalities that have become more common, and are often required, to interact with modern astronomical data interfaces.  Examples include items such as constant management of local paths to data files, learning how to write HTTP requests for accessing data served remotely, learning SQL to access data from databases, or even learning how to write web applications to serve data to others.  These kinds of tasks often end up as logistical overheads that can be frustrating for end users, as they take repeated time to learn or implement and become barriers.  These barriers can impede users' ability to do their science, which, at best, delays scientific discovery and, at worst, prevents accessing the necessary data altogether.

\begin{figure*}
\centering
\includegraphics[width=\textwidth]{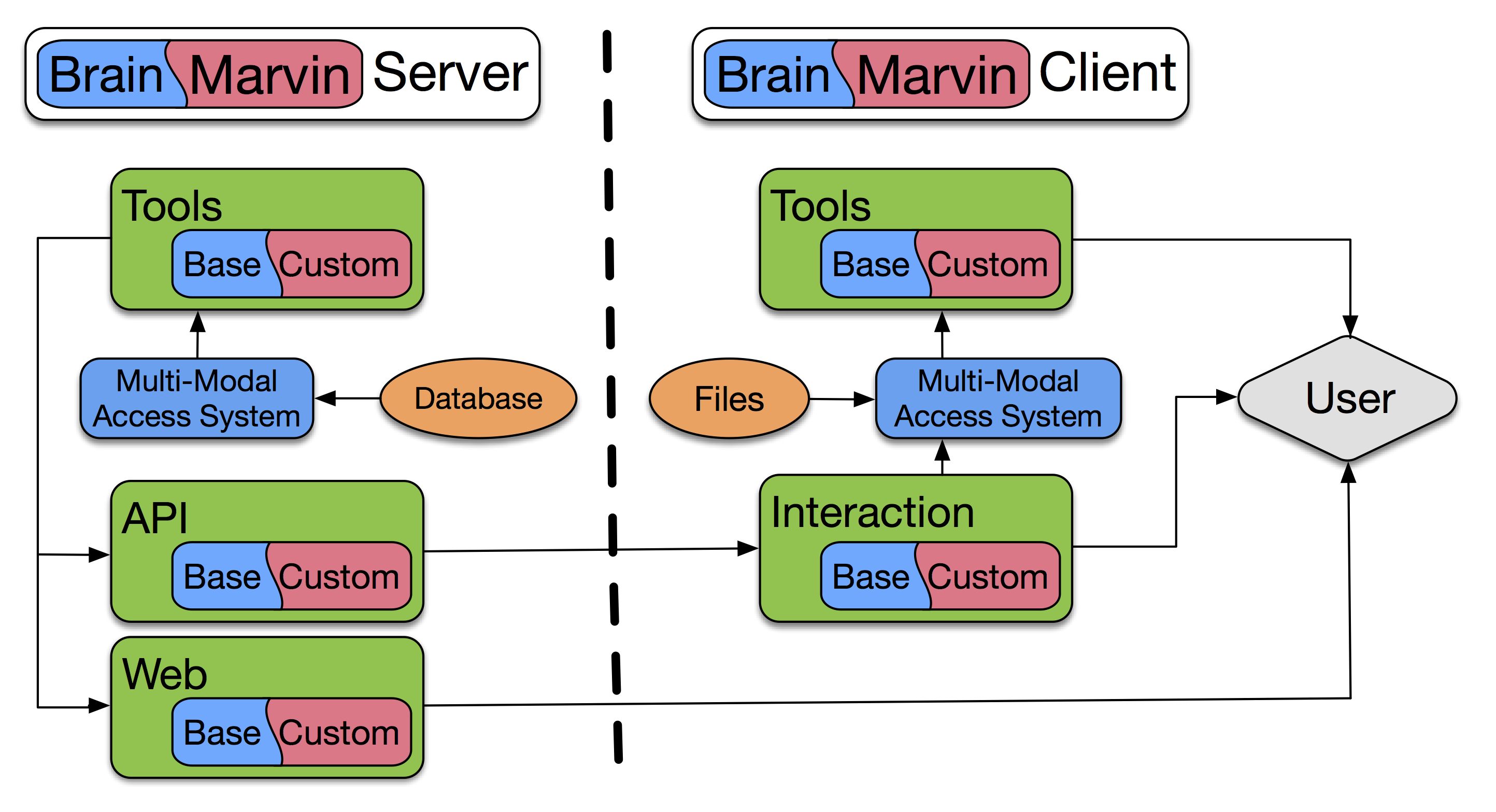}
\caption{\marvin\ data flow and the relationship between \marvin\ and the \brain.  Base classes in the \brain\ are subclassed into customized \marvin\ classes.  All data flows via the Multi-Modal Access system.  Client-side data flows through the Tools, and server-side data flows through the API or Web (through the Tools, \texttt{Interaction} class, or the browser).}
\label{fig:DataFlow}
\end{figure*}

The primary design goal with the development of \marvin\ was to abstract away these overheads, and provide a framework that automatically handles much of this management.  While \marvin\ is software specific to the MaNGA data set, many of these overheads are often independent of the type of data being served or accessed.   To facilitate easier access and potential reusability of these features for other projects, we have placed these kinds of features into an additional core product, called the \brain, which \marvin\ depends on.  Figure~\ref{fig:DataFlow} shows the relationship between \marvin\ and its \brain.  \marvin's \brain\ (shown in blue) exists as base classes that sit underneath all of the components within \marvin.  These classes act as templates that can be reused and customized for different applications.   Our aim is to continue to migrate existing common \marvin\ features into the \brain\ so others can utilize the same tools.

\subsection{\datamodel}
\marvin\ programmatically implements the unique MaNGA data model for each data release to abstract the Data Products for the MMA system and users.  The MMA system relies on the \datamodel\ to produce the same Data Product (e.g., \cube\ or \dmaps) from the correct data release regardless of whether it was instantiated from a FITS file, a database, or via the API.  This abstraction makes scientific reproducibility much easier.  It also enables users to programmatically navigate the Data Products without having to refer to the documentation.  \marvin\ simplifies the data model for users by utilizing \texttt{FuzzyWuzzy}, a fuzzy string matching algorithm, to fix incorrect but unambiguous user input (e.g., ``gflux ha'' maps to  ``emline\_gflux\_ha\_6564'').  The \datamodel\ is available as a standalone navigable object allowing access to the content and format of all MaNGA deliverables from a single location.  Additionally, individual data models are attached to every relevant \marvin\ Tool, providing an internal lookup that all Tools use for self-consistency, making them robust against any changes to the underlying data files.  As the format of the FITS files changes periodically between data releases, the structure of the Tools remains the same as the data model provides that intermediate go-between.  Finally, the documentation for the data model is automatically generated (see Section~\ref{sec:documentation}) for reference.

\section{Programmatic Tools}
\label{sec:marvin-tools}

\begin{figure*}
\centering
\includegraphics[width=\textwidth]{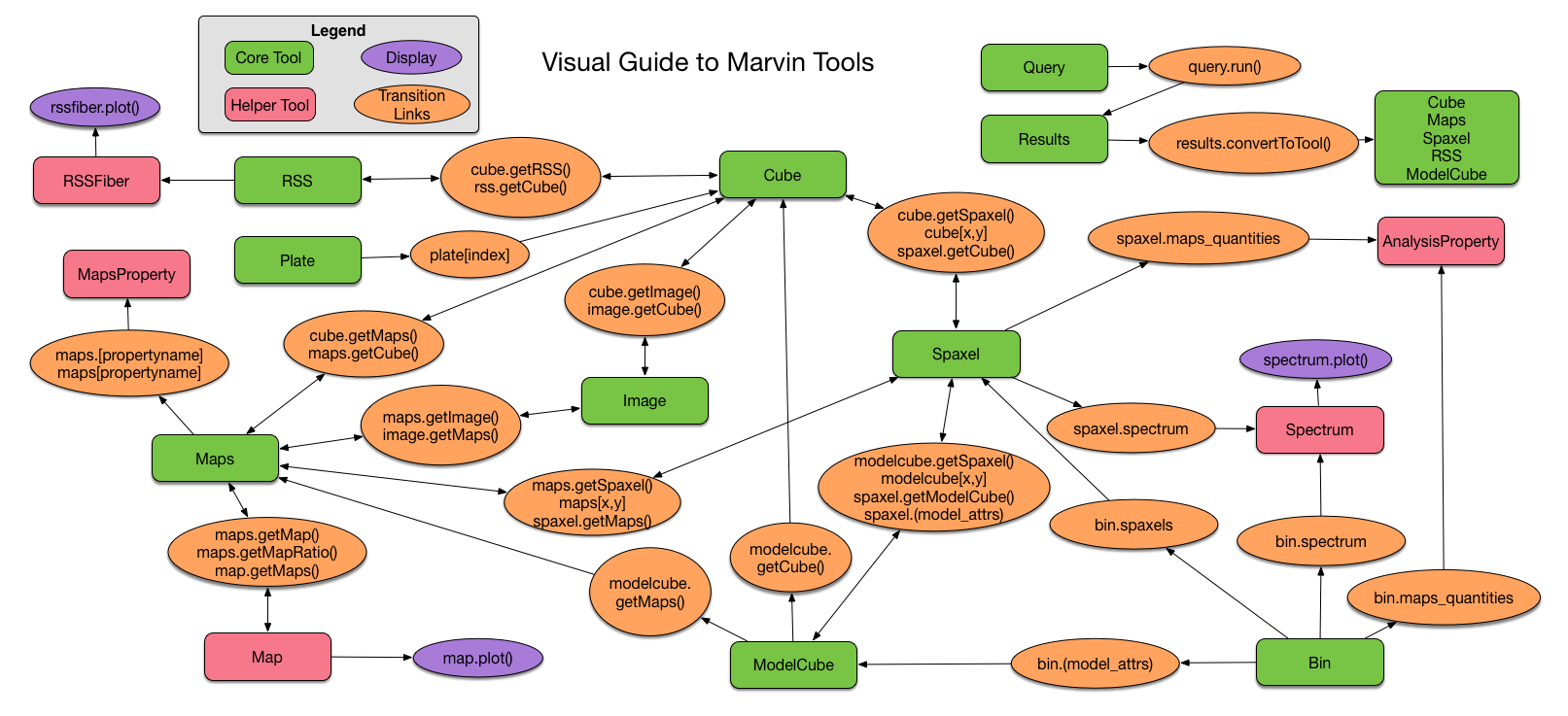}
\caption{A visual guide of the programmatic Tools, highlighting the complex interconnectivity between the tools.  Green icons represent core tool classes, with orange ovals showing the connections between them.  Pink icons are helper tools, and purple icons are endpoints for visually displaying data.}
\label{fig:VisGuide}
\end{figure*}

\marvin\ provides a programmatic interaction with the MaNGA data to enable rigorous and repeatable science-grade analyses beyond simply visualizing the data.  These tools come in the form of a Python package that provides convenience classes and functions that simplify the processes of searching, accessing, downloading, and interacting with MaNGA data, selecting a sample, running user-defined analysis code, and producing publication quality figures.  \marvin\ Tools are separated into two main categories: Data Product Tools and Query Tools.  The Data Product Tools are object-based and are constructed around classes that correspond to different levels of MaNGA data organization.  The Query Tools are search-based and are designed to provide the user the ability to remotely query the MaNGA galaxy data set and retrieve only the data they want.  \marvin\ also provides a built-in data model, which describes the science deliverables for every data release of \marvin.  Overall, these tools allow for easier access to the data without knowing much about the data model, by seamlessly connecting all the MaNGA data products, eliminating the need to micromanage a multitude of files.  Figure~\ref{fig:VisGuide} shows a visual guide to all our tools, and highlights the interconnectivity between them.

\subsection{Galaxy Tools}
\label{sec:galtools}

These tools cover four main classes, \cube, \rss, \dmaps, and \modelcube, that are associated with the analogous Data Reduction Pipeline (DRP; \citealt{law2016}) and Data Analysis Pipeline (DAP; Westfall et al.~in prep.) data products---namely, multi-dimensional data cubes, row-stacked spectra, derived analysis maps, and model cubes.   The four main tools all inherit from a common core object, thus sharing much of their functionality and logic, such as the MMA.   These tools are designed to do more than simply wrap and serve the underlying data and metadata contained in FITS files.  Their goal is to streamline the users' interaction with that data and simplify common but often non-trivial tasks associated with handling the data.  Via these tools, all data is delivered as \texttt{Astropy} \texttt{Quantity}s, with attached variance, mask, and any associated available properties.  With \texttt{Quantity} variance and mask tracking, this enables robust and consistent arithmetic between any of the DAP \dmaps.  Each tool has a built-in data model describing the format and content of the data it delivers.  This data model also provides convenient top-level access to all properties available, with autocomplete navigation.  Any given tool has convenient access to associated data products, as well as easy download capability for any data accessed remotely.

Features or functionalities that are common to multiple tools are designed as Python Mixin objects.  These objects are designed as isolated pieces of code that can be ``mixed in'' with any other tool, giving that tool access to its parameters.  Access to the NASA-Sloan Atlas (NSA) catalog \citep{blanton2011}\footnote{\url{https://www.sdss.org/dr15/manga/manga-target-selection/nsa/}} and the DAP summary file for instance are implemented in this manner.  Extracting spaxels within a specified aperture is a common functionality delivered to all tools as a Mixin.

There are additional tools that are not associated with a particular MaNGA data file but instead map to objects related to the MaNGA data.  These tools behave in much the same way as the core tools.  They utilize the MMA, allow for remote file downloading, and are seamlessly integrated with each other.  The \plate\ tool corresponds to an observed SDSS plate used during MaNGA observations.  This object provides a list of all of the {\cube}s observed on a given plate, along with additional metadata associated with the plate, e.g., exposure numbers, observation date, etc.  The \image\ tool provides interaction with the MaNGA IFU image cutout from SDSS multi-band imaging.  It allows for quick display of the IFU image, over-plotting of the IFU hexagon, over-plotting of the individual IFU bundle or sky fibers, or generating an entirely new image at a custom pixel scale.  Additionally a list of \image\ objects can be quickly generated and downloaded to the local client system.  Image utilities also exist to quickly download a list of images in bulk using the streaming capability of \sdssaccess.

\subsection{Sub-Region Tools}

Marvin provides sub-region galaxy tools, which are designed to access individual components within the main MaNGA data products.  \rssfiber, \spaxel, and \bin\ provide access to the row-stacked spectra from individual fibers, datacube spaxels, or bins (for binned DAP data), respectively.  These tools come with convenient plotting functions, as well as access to all the DRP and DAP properties associated with a given element.  The DAP produces data products with different spectral binning schemes for different science cases: unbinned spectra (SPX), spectra binned to S/N$\sim$10 using the Voronoi binning algorithm (VOR10), and a hybrid binning scheme (HYB10), with spectra binned to S/N$\sim$10 for the stellar kinematics, but emission-line measurements are performed on the individual spaxels.  The ``HYB10'' binning type for DAP products has complicated the underlying binning scheme of spaxels.  The \spaxel\ and \bin\ tools make the binning much more straightforward.  Each \spaxel\ property contains information about whether it is binned or not, hooking into the \bin\ tool when appropriate.  The \bin\ tool displays only the relevant information for the underlying property and binning type, clearing up most of the obfuscation with accessing the ``HYB10'' binned files directly.  From \bin, one can access all spaxels belonging to that bin, as well as generate masks for that bin.

\subsection{Query Tools}
\label{sec:query}

\marvin\ provides tools for searching the MaNGA data set through an SQL-like interface, either via a web-form or a Python class.  The \marvin\ \query\ system uses a simplified SQL syntax (see Section \ref{sec:pnls}) that focuses only on a filter condition using boolean logic operators and a list of parameters to return.  Not only does this simplify the syntax, but it automatically performs the incredibly complex table joins required to extract data from the MaNGA database.  Users can query the MaNGA sample on global galaxy properties, similar to searching through the DRP and DAP summary files.  In the near future, users will be able to perform intra-galaxy queries on individual spaxel measurements---a task that requires a database or loading all of the MaNGA spaxel data into RAM.  Tutorials for querying with \marvin\ are available in the online documentation\footnote{\url{https://sdss-marvin.readthedocs.io/en/stable/query.html}}.

\subsubsection{Pseudo-natural Language Syntax}
\label{sec:pnls}

\begin{figure*}
\centering
\includegraphics[width=\textwidth]{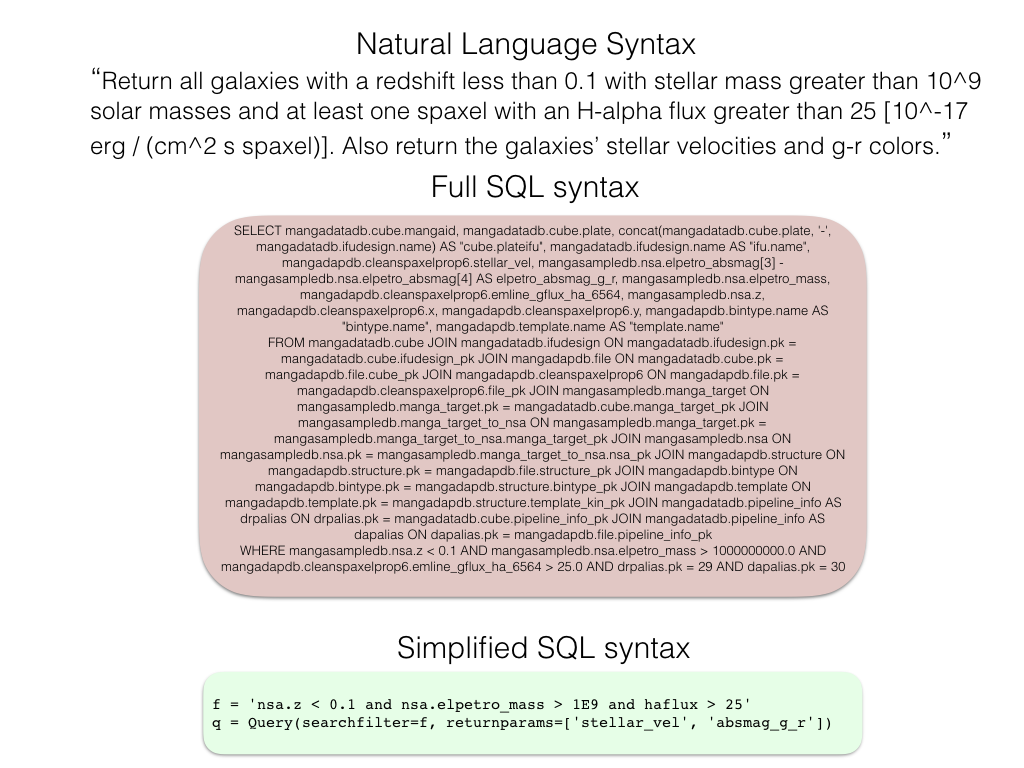}
\caption{An example query on the MaNGA data set.  The top describes the query in natural language syntax, i.e., how a user would describe it.  The red panel shows the full SQL syntax needed to perform the same query on the MaNGA database.  The green panel shows the corresponding pseudo-natural language syntax, and its use in \marvin.}
\label{fig:ExQuery}
\end{figure*}

Figure~\ref{fig:ExQuery} shows an example of a MaNGA query in (1) natural language, (2) full SQL syntax, and (3) simplified pseudo-natural language syntax.  While the query is relatively easy to describe in natural language, the full SQL syntax (red panel in Figure~\ref{fig:ExQuery}) is immensely complicated  to construct, even if the user already knows how to write SQL queries.  SQL queries consist of three main parts: a \texttt{select} clause, a \texttt{join} clause, and a \texttt{where} clause.  Constructing the \texttt{select} and \texttt{join} clauses require detailed knowledge of the MaNGA database schema, table design, available columns, and the keys needed to join the tables.  With the \marvin\ \query\ tool, rather than submitting the full SQL query, the user submits only a simplified \texttt{where} clause and an optional list of properties to return.  The remainder of the query (the \texttt{select} and \texttt{join} clauses) is built dynamically behind the scenes, converted to raw SQL, and then submitted to the database.  This allows the user to focus on the properties and their values in the selection criteria.

\marvin\ uses SQLAlchemy to map Python ``model'' classes onto each of our database tables and columns.  This provides the base ability to dynamically build and submit SQL queries in Python.  With these model classes, \marvin\ constructs a singular look-up dictionary containing a mapping between a string parameter name, in the form of \texttt{schema.table.column\_name}, and its Python counterpart.  This provides an automatic way of looking up the database location for a given parameter name, effectively removing the \texttt{select} clause.   \marvin\ uses \texttt{networkx} to map those model classes onto a network tree, which allows the construction of a proper SQL \texttt{join} clause given any two input parameters across all tables in all schema in the database.  Finally, \marvin\ uses a customized version of \texttt{sqlalchemy-boolean-search} to simplify the \texttt{where} clause to a simple input string.  This is a boolean parser which takes a string boolean filter condition, parses it, and converts to the proper SQLAlchemy filter object.  The green panel in Figure~\ref{fig:ExQuery} shows the pseudo-natural language equivalent of the desired query.

\subsection{Utilities}
\label{sec:utilities}

\textbf{Maskbit}: MaNGA uses masks as a compact way to simultaneously convey information about the status of an object under many boolean conditions.  The MaNGA pipelines produce quality masks at each processing stage, which allow users to filter out specific types of undesirable data when performing science analyses.  During target selection, MaNGA likewise creates targeting masks that encode the sample or program under which an object was selected to be targeted.

The Maskbit class is a general purpose utility used by other Data Product Tools.  It automatically loads the schema for a mask, which can be easily displayed for the user.  It can then convert between the native integer value (e.g., 1025) to the list of bits set (e.g., [0, 10]) to the corresponding list of labels (e.g., [``NOCOV'', ``DONOTUSE''] for the \texttt{MANGA\_DRP3PIXMASK} mask, which indicates that the spaxel has no coverage in the cube and should not be used for science).  Users can create a mask by providing a list of labels instead of filtering bits.  This class also enables searching on bits, which is particularly useful for target selection using the targeting masks.

\textbf{Plotting Utilities}: \marvin's plotting utilities enable users to quickly display images, spectra, and maps of individual MaNGA galaxies or galaxy sub-regions.  The plotting utilities also can put galaxies or sub-regions in context via scatter and histogram plots of query results.  Beyond the image utilities, which have been described previously, and the spectrum plotting, which is a straightforward line plot, we will describe the map, scatter, and histogram plotting in more detail below.  As a general philosophy, \marvin's plotting utilities are designed to have smart defaults for quickly making useful visualizations while allowing for significant customization via standard \texttt{Matplotlib} methods.

One of the most difficult aspects of generalized map plotting is automatically setting the range of the color bar without being overly sensitive to poor measurements or outlier values.  Map plotting automatically masks spaxels with poor measurements as flagged by the DAP or due to low signal-to-noise ratios.  Users can tailor the masking by specifying flags, creating new masks, or changing the minimum signal-to-noise ratio.  These masked regions are differentiated from areas outside of the IFU footprint to distinguish between regions with poor measurements and regions without data.  To handle good but outlying measurements, \marvin's map plotting does percentile clipping by default but allows for sigma clipping or a user-defined range.  There also is a logarithmic option to help display properties with large dynamic ranges.  It automatically uses a symmetric color bar for velocity maps since there is a natural zero point.  \marvin\ also has an option for creating discrete color maps to show, for instance, spatial regions whose nebular line ratios are consistent with photoionization via star formation or a central AGN.

\begin{figure}
     \subfloat[Matplotlib scatter plot for results with less than 1000 points.\label{subfig:scatter}]{%
       \includegraphics[width=0.45\textwidth]{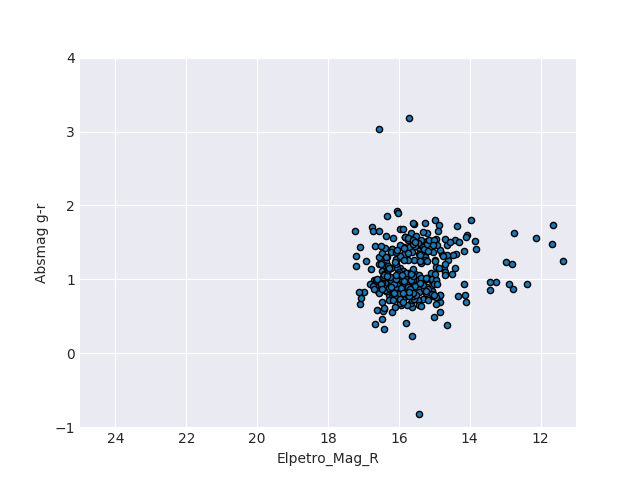}
     }
     \hfill
     \subfloat[Matplotlib hexbin plot for results with between 1000-500,000 points.\label{subfig:hex}]{%
       \includegraphics[width=0.45\textwidth]{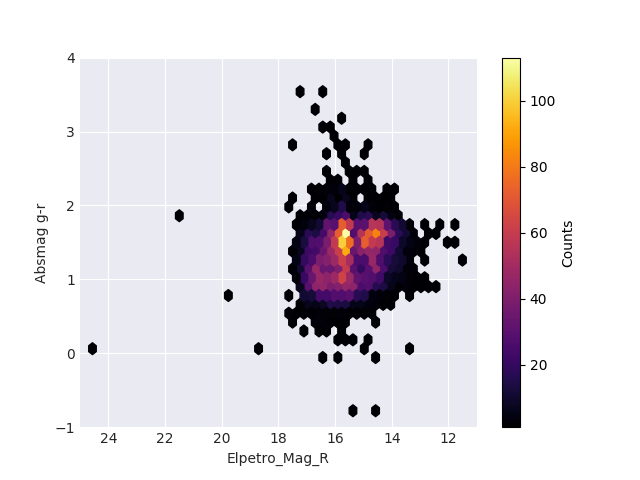}
     }
     \hfill
     \subfloat[Matplotlib scatter density plot using \url{https://github.com/astrofrog/mpl-scatter-density} for results with more than 500,000 points.\label{subfig:density}]{%
       \includegraphics[width=0.45\textwidth]{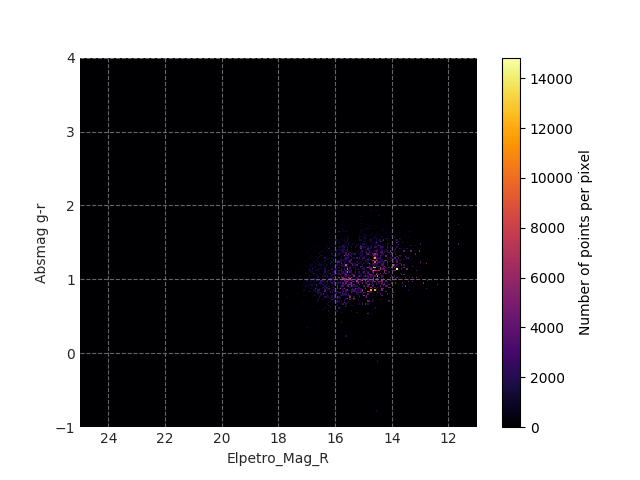}
     }
     \caption{Scatter plotting capability from the \marvin\ \results\ Tool.  Depending on the number of results, \marvin\ plots using a straight scatter plot, a hex-binned density distribution, or a scatter density map.  Note: this figure uses the same scatter plot in each subfigure to illustrate the differences in plotting styles rather than separate plots with more accurate result counts.}
     \label{fig:QueryPlots}
   \end{figure}

Querying is one of \marvin's most powerful features.  Yet it is difficult, if not impossible, to discover trends in large tables of text produced from a query.  To that end, \marvin\ includes utilities to make scatter and histogram plots of query results.  Queries in \marvin\ can return results with anywhere from a few to millions of data points, so \marvin's scatter plot changes the underlying display technique depending on the number of data points (see Figure~\ref{fig:QueryPlots}).  Fewer than 1,000 data points are shown individually (\ref{subfig:scatter}), 1,000--500,000 data points are shown as a hex binned density distribution (\ref{subfig:hex}), and more than 500,000 data points are shown as a scatter density map (\ref{subfig:density}) that is responsive even with millions of data points.  By default, scatter plots show marginal histograms with the mean and standard deviation.  Users can also create histograms separately from a scatter plot and extract the data points in each bin.

\textbf{Analysis Tools}: At the time of publication, we have prioritized development of aspects of \marvin\ required for interfacing with the MaNGA data over providing downstream analysis tools.  However, \marvin\ is ideally suited to serve as a foundation for analysis tools that extend its functionality to additional processing steps.  One such analysis tool that has already been developed is a tool to classify different regions of a galaxy according to classical emission line ratios.

As discussed by, e.g. \citet{baldwin1981} (BPT hereafter), the nebular permitted (\Ha, \Hb) and forbidden emission line transitions (e.g., \otwo\ $\lambda 3727$, \othree\ $\lambda 5008$, \ntwo\ $\lambda 6585$) are commonly strong and easy to detect in galaxies that contain significant quantities of gas.  Since these transitions have different ionization potentials, their relative strengths encode information about both the metallicity of the gas and the hardness of the radiation field emitted by the source of ionizing photons.  As such, easily-measured line ratios such as \othree/\Hb\ and \ntwo/\Ha\ can be used to discriminate between \htwo\ regions produced by thermal (i.e., star formation) and non-thermal processes (e.g., shocks and active galactic nuclei).

\marvin's BPT tool returns masks in which individual spaxels have been classified as ``star-forming'', ``Seyfert'' or ``LINER-like'' line ratios, such that a user can then plot diagnostic diagrams, categorial maps of the classifications, or maps filtered by these classifications (for instance, plotting the \Ha\ flux for star-forming regions).  Such analyses have revealed significant clues as to the physical origins of the ionizing photons, indicating for example that in many sources observed to have LINER-like line ratios in SDSS single-fiber spectroscopy the gas is spatially extended and likely ionized by hot evolved stars rather than a central AGN \citep{belfiore2016}.

\subsection{Contributed Code}
\label{sec:contribcode}

While the core of SDSS data releases centers around its base projects' science deliverables, smaller teams frequently contribute added value to its core deliverables with additional science products.  These value-added data products or catalogs (VACS) are derived data products based on the core deliverables that are vetted, hosted, and released by SDSS in order to maximize the impact of SDSS data sets.  To increase the visibility of MaNGA VACs, \marvin\ has hooks to allow users to contribute small pieces of code that plug their VACs into the overall system, immediately connecting their VAC into the larger suite of \marvin\ Tools and MaNGA Data Products.  Each contributed code piece is well-documented, adheres to the overall standards set by SDSS, and contains the proper software credit for the user.

The core design principles of \marvin\ are to perform most of the legwork for the users, making access as easy as possible, while allowing users to contribute their own code to help expand \marvin's functionalities.  For VACs, contributors create a new class defining their VAC based on a pre-defined base class which provides unique target identifiers and automatic file retrieval methods needed to extract specific data from files.  Contributors simply define the name of their VAC, the unique file path parameters, and a single method returning the data content.  Contributors do not need to implement access to the core data products, which is already handled by \marvin.

More generally, the \marvin\ code is structured to ease contributions of drop-in utility or analysis methods that add functionality to \marvin.  These functions can manipulate or extract data from existing \marvin\ Tools, perform some analysis, or return a plot or data.  The BPT tool from the previous section serves as an example of such a drop-in function that easily wraps the existing Tools.  Users are encouraged to contribute \marvin-based analysis code back into the project so that others can take advantage of their efforts.

\section{Marvin Web}
\label{sec:marvin-web}

The web, or browser-based information gathering, is often the first entry point for any user new to a field.  Poor web design (e.g. cluttered content, complex interfaces) can quickly discourage users from interacting with the delivered content.   \marvin\ provides a web front-end that aims to be as intuitive and streamlined as possible, with a focus on quick visual exploration of the MaNGA data set, leaving more rigorous analysis to the programmatic \marvin\ Tools or the user's own scripts.  This minimal but interactive interface encourages users to quickly engage with MaNGA data and, when ready, seamlessly transition into more advanced environments.   Our web component is built using \flask, a Python-based, lightweight, micro web-framework.  \flask\ allows for quick deployment of a web application with minimal effort.  It contains its own built-in web server for small scale deployment, or can easily be integrated into more advanced web-servers for production deployment.  It has built-in hooks for modularity and extensibility, and employs a templating system for writing front-end code like HTML or Javascript in a modular way.

\marvin\ Web currently provides the following features:
\begin{itemize}
\item a Galaxy page, for detailed information and interaction with individual galaxies in MaNGA,
\item a Query page, for searching the MaNGA data set using the simplified SQL pseudo-natural language syntax described in Section \ref{sec:pnls},
\item a Plate page, containing all MaNGA galaxies observed on a given SDSS plate, and
\item an Image Roulette page that randomly samples images of MaNGA galaxies, useful for browsing the wealth of variety in the 10,000 galaxy sample.
\end{itemize}

\begin{figure*}
\centering
\includegraphics[width=\textwidth]{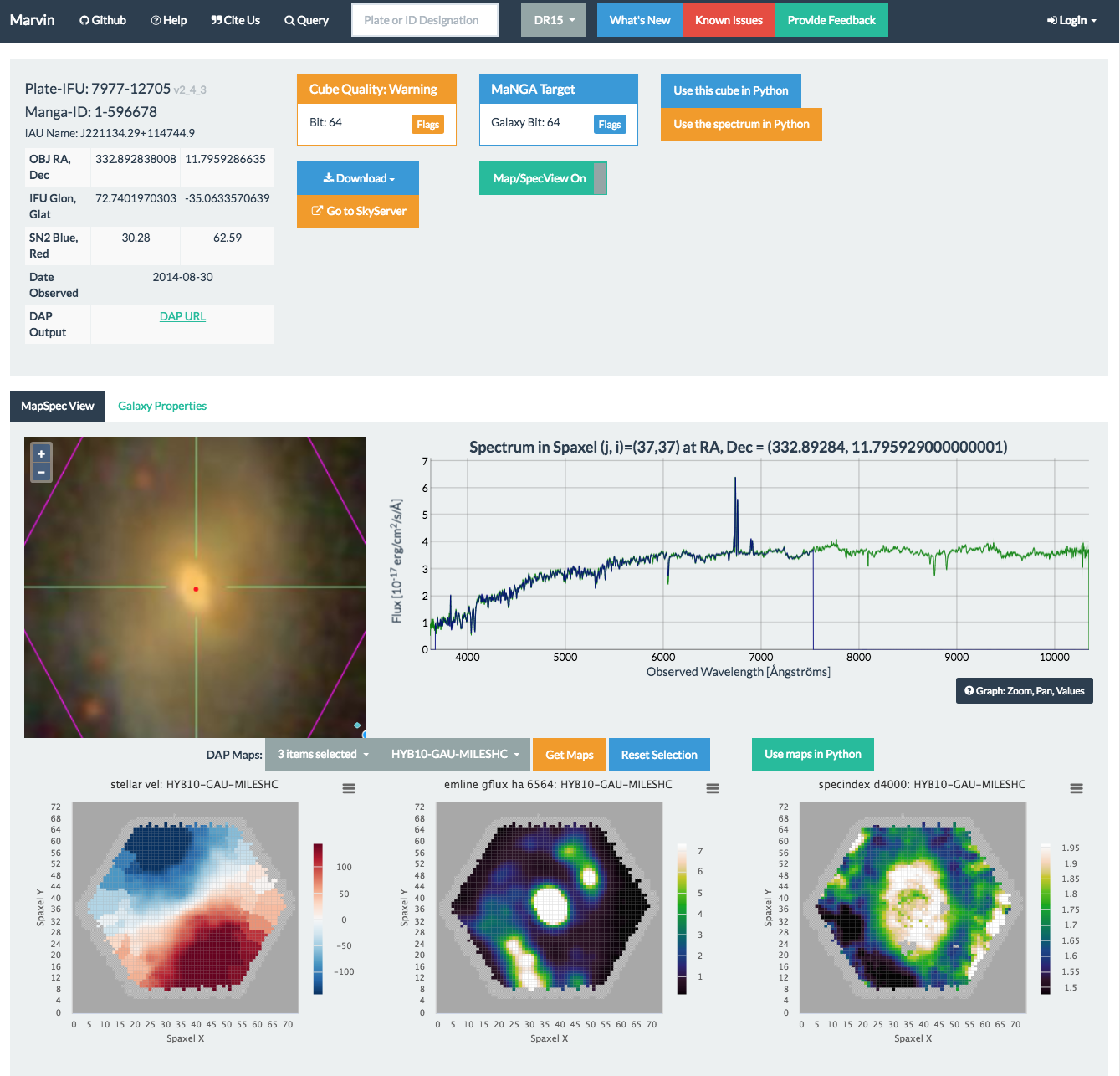}
\caption{The \marvin\ web Galaxy page, highlighting the interactive point-and-click feature.  Users can dynamically interact with individual galaxy spaxels and maps.  Clicking anywhere within the optical image, or the DAP maps, retrieves the spectrum at that spaxel location. The interactive spectrum displays both the flux (green) and model fit (blue) for the selected spaxel.  Marvin displays three maps by default: the \Ha\ emission line flux, the stellar velocity and D4000 spectral index.  Dropdown menus provide additional maps to display or maps of different binning schemes.}
\label{fig:MarvinWeb}
\end{figure*}

The Galaxy page (see Figure~\ref{fig:MarvinWeb}) provides dynamic, interactive, point-and-click views of individual galaxies to explore the output from the MaNGA DRP and DAP, namely, spectra and map properties, along with galaxy information from the NSA catalog.  In contrast to the prototype, this page is completely interactive, with more galaxy metadata.  These interactive features are deployed using a variety of third-party Javascript libraries:  \texttt{Dygraphs} for the spectral viewer,  \texttt{Highcharts} for the map and scatter plot viewers,  \texttt{OpenLayers} for the interactive optical image display, and  \texttt{D3} for the box-and-whisker plots.

\begin{figure*}
\centering
\includegraphics[width=\textwidth]{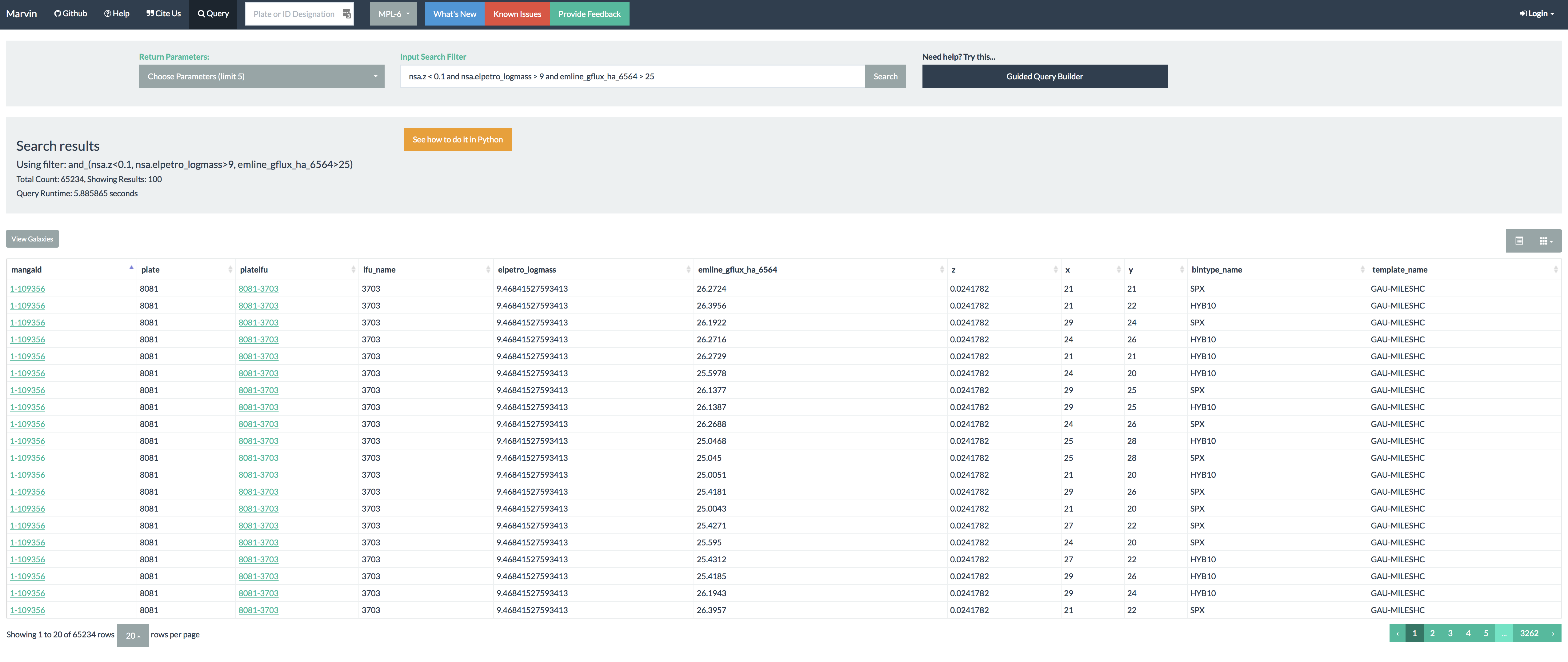}
\caption{The \marvin\ web Query page, highlighting an example query using the simplified \query\ tool syntax, presents the results in a navigable table. The basic Query page components consist of a simple text box for inputting a search filter, and a drop-down menu of a subset of available parameters to return in the query.  Additionally there is a button to construct a search filter in a guided fashion.}
\label{fig:web-query}
\end{figure*}

The Query page (Figure~\ref{fig:web-query}) provides the entry point for quickly searching through the MaNGA dataset.  In contrast to the prototype, this page provides search capability for all properties in the DRP and DAP summary files, with minimal impact on the design interface of the page. The search capability will soon be extended to the entire suite of MaNGA parameters.  It is built on top of the \marvin\ \query\ tool, providing a single simple interface, for both the web and tool, that one needs to become familiar with.  In addition, the query system can be easily extended for both web and client users at the same time.  Performing a query produces a navigable table of results, with each row linking to the individual galaxy.  One can optionally switch to a postage stamp view of all galaxies returned in the query subset.

\begin{figure*}
\centering
\includegraphics[width=\textwidth]{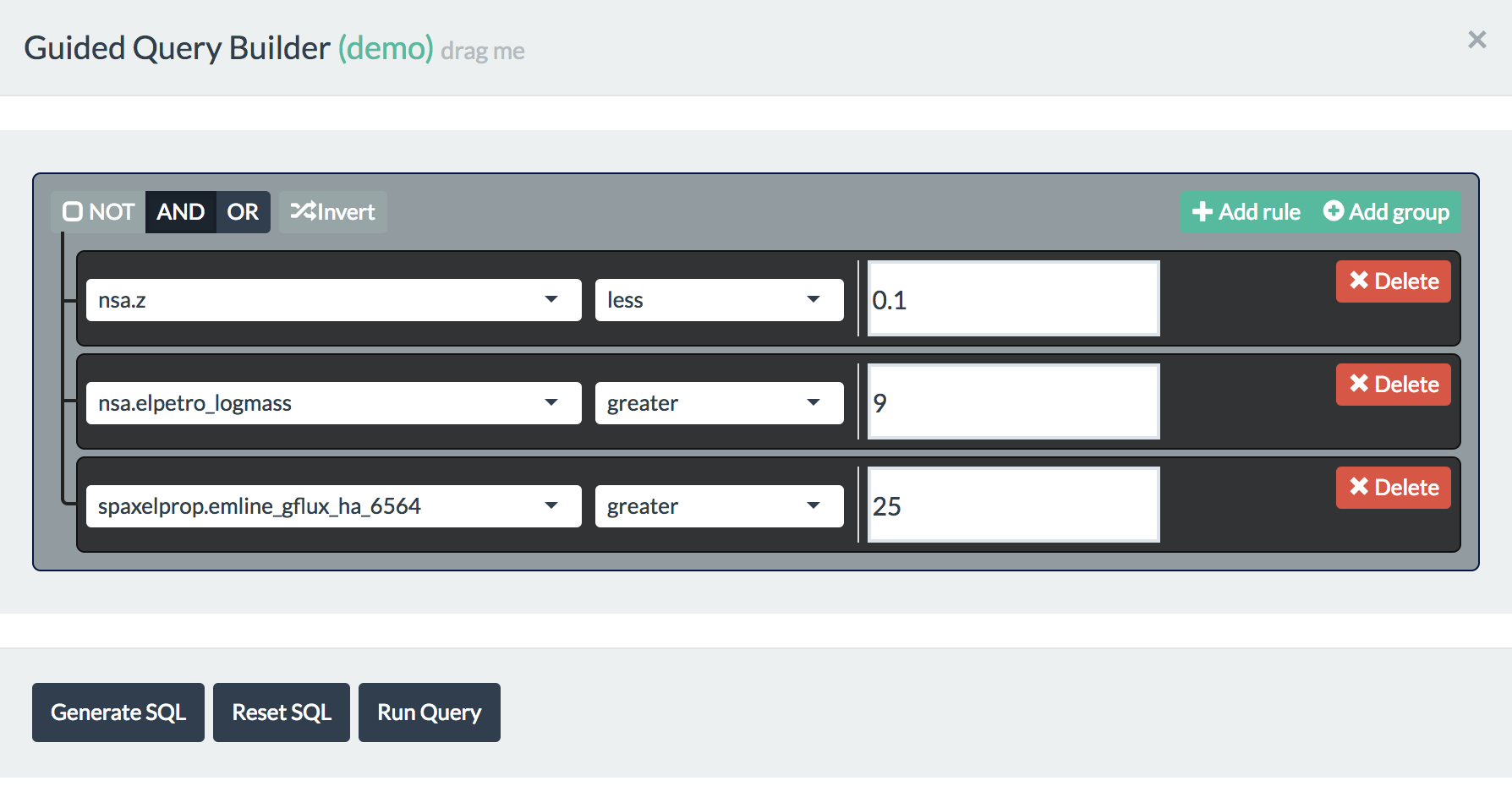}
\caption{A guided SQL builder for constructing complex queries on the \marvin\ web Query page.  This query maps to the search ``return all galaxies with a redshift less than 0.1, a stellar mass greater than $10^9$ solar masses, and at least one spaxel with \Ha\ flux greater than 25~$10^{-17} erg/cm^2~s~spaxel$". }
\label{fig:web-guidedsql}
\end{figure*}

While the input structure to the \marvin\ \query\ tool is simplified greatly from the underlying full SQL statement, the syntax can still be complicated to learn.  Some users may find it cumbersome, delivering confusion instead of intuition.  The Query page also includes an interface for dynamically constructing a SQL statement in a guided manner (see Figure~\ref{fig:web-guidedsql}).  This interface provides a series of parameter drop-downs which, in conjunction with operators and values, can be used to build conditions, and combined together with boolean operands.   A web video tutorial\footnote{\url{https://www.sdss.org/dr15/manga/manga-tutorials/marvin-tutorial/marvin-web/}} is available highlighting general usage, with more information available in the online documentation.

Because the web components are built on top of the Tools, all the galaxy and query features can be mapped to an underlying equivalent \marvin\ tool command.  This allows users to recreate what they experience in the Web with the \marvin\ Tools locally on their system.  On each page we provide feature-specific code snippets that indicate the equivalent commands for viewing galaxy maps or spaxels, or querying the dataset.  These snippets can be copied and pasted directly into the local iPython terminal.

For the back-end, \flask\ provides the basis for the framework as well.  It can be run in a ``debug'' mode for rapid development or be served in a production environment.  For production deployment, \marvin\ is run using the NGINX web-server, with uWSGI acting as the gateway interface between the \flask\ web-app and NGINX. \flask\ provides the basic framework on which the \marvin\ API and the web-facing front-end is built.  Our back-end \flask\ routes are built using the same suite of \marvin\ Tools available to the user on the client side.  In this manner, we can build a single tool for the user while also using it to provide the same content directly over the API or integrated into the front-end as a web features.  The server-side \marvin\ uses the same MMA system to determine data location, pulling first from a local database hosted at Utah, then from the files located on the Science Archive Server (SAS) filesystem.

\section{Marvin Back-End}
\label{sec:marvin-backend}

\subsection{REST-like API}
\label{api}

\begin{figure*}
\centering
\includegraphics[width=\textwidth]{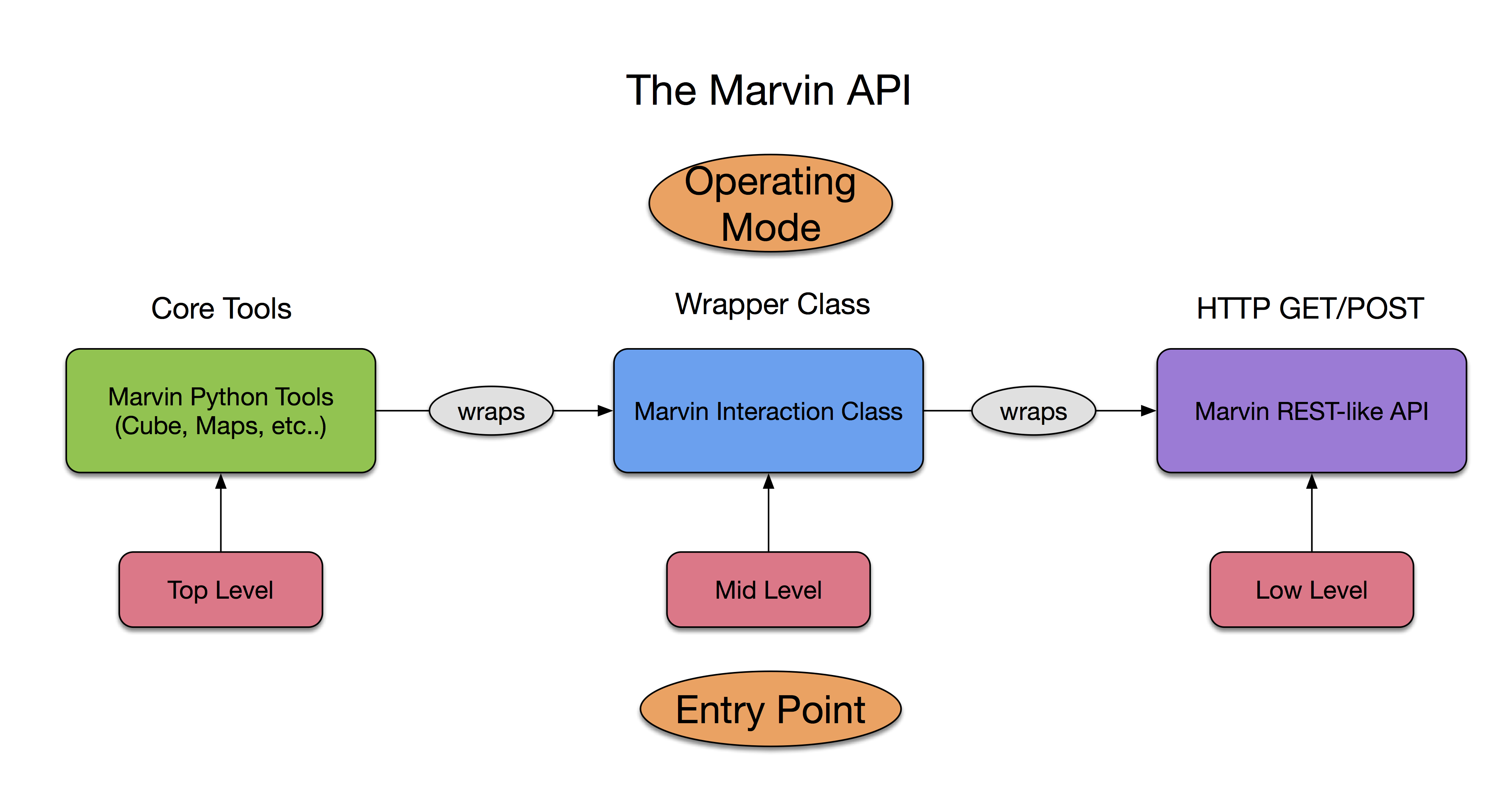}
\caption{The three layers of the \marvin\ API highlighting the process of a remote Tool call.  The top level (left) sits the \marvin\ Tools and is the main entry point for users.  The middle layer (middle) provides a wrapper class for handling request logistics.  The low layer (right) performs the standard HTTP requests.  Users naturally start at the top-most layer, but may use any layer for performing remote requests.}
\label{fig:MarvinAPI}
\end{figure*}

To provide remote data access, Marvin employs a REST-like web API, which defines a set of rules for remote data acquisition through HTTP request methods (i.e., \texttt{GET} and \texttt{POST}).  The API handles all requests and responses between the user and server.  There are three ways to interface with the \marvin\ API: directly through HTTP (low level), with a Python helper class (mid level), or via Marvin Tools (high level) (see Figure~\ref{fig:MarvinAPI}).

The lowest access level provides direct HTTP access to the API routes.  Our API routes use the underlying \marvin\ Tools and provide remote access to the most commonly desired features of the MaNGA data set.  A list of the available routes, and what data they provide, can be found in the online documentation\footnote{\url{https://sdss-marvin.readthedocs.io/en/stable/api/web.html}}.

While an experienced user can directly use the HTTP routes to retrieve data, not everyone is familiar with how to handle HTTP requests and responses.  The middle layer wraps the direct API calls into a Python \texttt{Interaction} class.  The \texttt{Interaction} class utilizes the \texttt{requests} package to handle \texttt{GET}/\texttt{POST} exception handling, set default request parameters, check the response, and provide convenience methods for parsing return data into Python data-types.

At the highest level, the \texttt{Interaction} class is built into the core \marvin\ Tools (and any Tools customized from them), providing remote access ability to all Tools.  \marvin\ contains a lookup dictionary for resolving API URL shortcuts into their full route names.  This dictionary allows each Tool to understand its required remote call and provides robustness against server-side API route changes.  In this access level, the user takes a hands-off approach to API requests.  The Tool determines when to access data locally or remotely.  If remote data access is required, it performs the API request without user input and shapes the data properly upon receiving the response.  When using the API, the Tools employ a lazy-loading approach to remote data to minimize server load.  The API returns the minimum amount of information needed to satisfy the user's request.  Additional information requested through the Tools is acquired through additional API calls.

\subsection{Database}

All MaNGA data is stored in a PostgreSQL relational database.  The database is the bedrock data storage component of \marvin.  It provides the basis for interactive visualization in the web, spatially-resolved queries, and selective data retrieval.   For each release of MaNGA, we store the metadata and raw spectral output from the DRP and DAP pipelines.  We currently store data from MaNGA internal data releases, referred to as MaNGA Product Launches, MPLs, 4-7, and the public data release DR15.   The current size of the MaNGA database is 9~TB for 20153 galaxies across MPLs 4-7.  The database contains three main schema: \texttt{mangadatadb} which contains the DRP output; \texttt{mangadapdb} which contains the DAP output; and \texttt{mangasampledb} which contains information on MaNGA targets and the NSA catalog.  In addition, there is an auxiliary schema for miscellaneous data and a history schema which stores user and query metrics.  The main schema designs can be found in the online documentation\footnote{\url{https://sdss-marvin.readthedocs.io/en/stable/api/db.html}}.

\subsection{SAS Filesystem}

The SAS filesystem is the data warehouse for SDSS.  Hosted at the University of Utah and mirrored at the National Energy Research Scientific Computing Center, the SAS includes all of the raw and reduced SDSS data, including the intermediate and final data products from each survey's data reduction or analysis pipelines and value-added catalogs ($\sim$1000~TB).  The SAS serves both the collaboration (through a private gateway) and the public.  The filesystem structure is organized hierarchically by survey to facilitate easy navigation.  Both the software and data products are under version control, and the versions are explicitly included in the file paths.  The explicit versioning in the file paths allows for consistent access and rapid deployment for internal and public data releases.  All software and data products are frozen on a schedule set by the SDSS collaboration and tagged to maintain a self-consistent, reliable, and robust data system.  These immutable tags and frozen reduction versions ensure the reproducibility of high-quality products.  By developing \marvin\ to work on top of this structure, we can consistently deliver an archive-quality data product to the community, mitigating concerns about underlying intermittent data changes.

\subsection{Authentication and Access}
\label{sec:auth}

\marvin\ provides access to MaNGA data for both the SDSS collaboration and the public astronomy community.  The SDSS collaboration provides data access rights for a proprietary period, so \marvin\ has collaboration-only and public access modes.  The collaboration-only access mode provide access to the private gateway for both internal data releases and public data releases.  In contrast, the public access mode provides access to the public gateway for only the public data releases.  Approved SDSS collaboration members must authenticate with \marvin\ before access is granted to the private gateway. For the Web, \marvin\ uses the \flask\ extension \texttt{Flask-Login}, which uses session-based cookies to handle all login and authentication.  For the API, \marvin\ uses \texttt{Flask-JWT-Extended}, which authenticates via JSON Web Tokens.  After supplying their credentials in a Unix standard \texttt{netrc} file, the user is allowed to login and receive a valid token.  The token is inserted into every API request and authenticated on the back-end.

\section{Workflow}
\label{sec:usecase}

\begin{figure*}
\centering
\includegraphics[width=\textwidth]{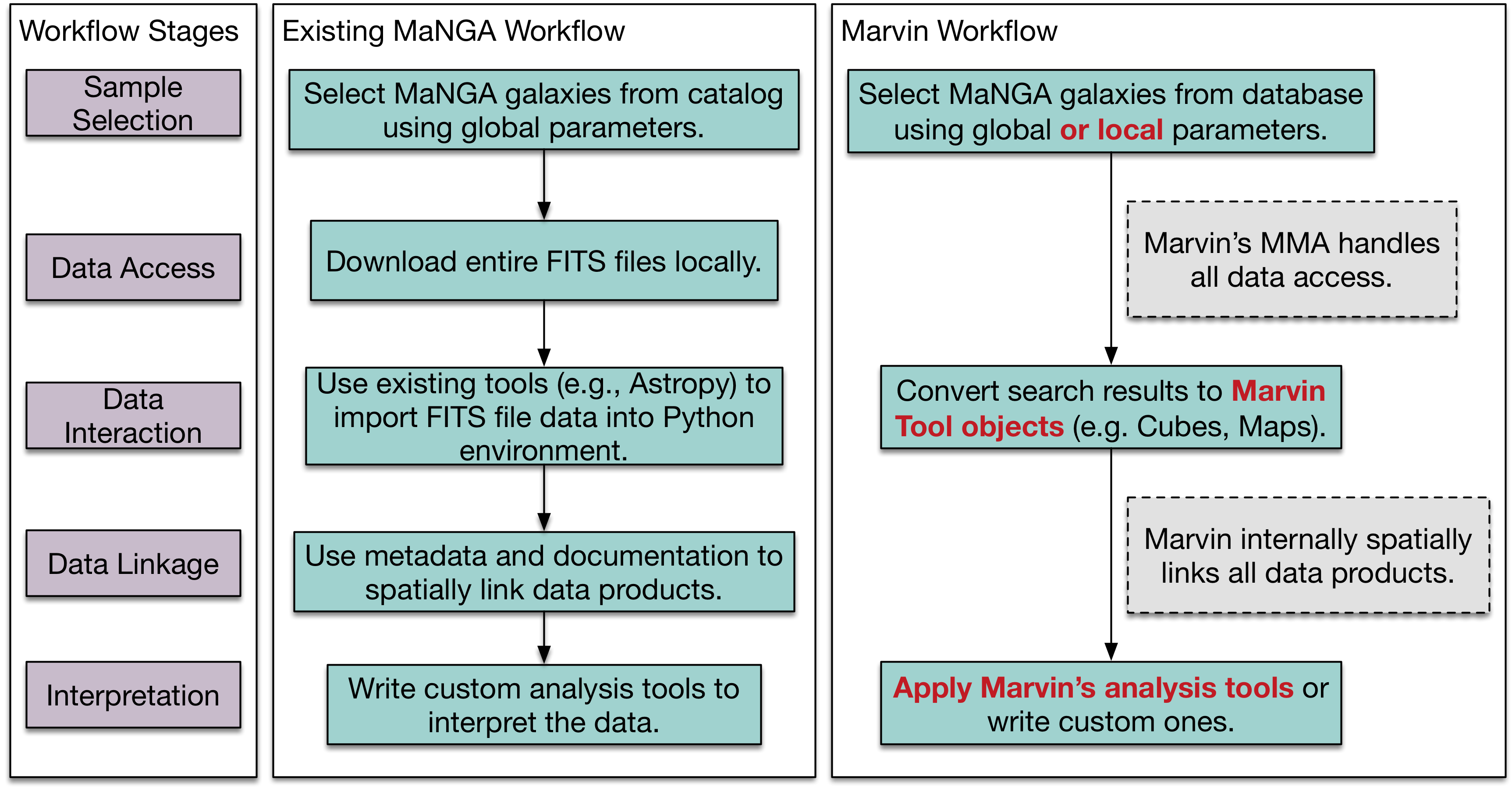}
\caption{Typical MaNGA workflow stages from sample selection through interpretation (left column) performed with existing tools (middle column) and \marvin\ (right column).  At each stage, \marvin\ either enhances the existing capabilities (highlighted in red) or automatically handles tasks (gray boxes).}
\label{fig:Workflow}
\end{figure*}

Figure~\ref{fig:Workflow} highlights an example workflow going through the stages of Sample Selection, Data Access, Data Interaction, Data Linkage, and Interpretation.  Prior to \marvin, the workflow for an analysis of MaNGA data, as indicated in the middle panel, would typically consist of a user selecting galaxies based on global galaxy parameters, downloading all the data files for those galaxies locally, then using existing tools (e.g., Astropy) to load the data into generic (i.e., not MaNGA-specific) FITS objects in a programming environment.  To retrieve all relevant information for a target, the user must load, access, and construct the spatial links between data in separate files.  Finally, users must write their own custom, often reinvented, analysis tools to visualize and interpret the data for their science.

The \marvin\ framework streamlines the existing workflow as shown in the rightmost panel of Figure~\ref{fig:Workflow}.  \marvin\ enhances existing workflow steps (shown as red text) and obviates  workflow steps that require logistical overhead effort for data handling (shown in gray dashed boxes).  While the methods involved in the existing workflow are functional, they contain the following problems, which \marvin\ redresses:

\begin{itemize}

    \item The first four tasks in the workflow, i.e. selecting a target sample, downloading and linking FITS files, are easy to describe but require moderate effort to implement, which is compounded when iterating over selection criteria during exploratory analyses.  In contrast, \marvin\ provides functionality to handle these ubiquitous tasks.  In particular, \marvin's front-end web interface enables rapid preliminary visual exploration without downloading data or writing code.

    \item The selection of galaxies to analyze can only be done on global quantities, not the maps nor the spectra.  In contrast, search capabilities in \marvin\ are far more powerful, flexible, and detailed.  \marvin\ can perform complex queries on the maps and spectra (e.g., search for galaxies with a high star formation rate surface density near the center).

    \item Unnecessary data is inevitably downloaded, since only entire files (containing the entire data cube or hundreds of maps) are available for download, increasing bandwidth and disk resources.  In contrast, \marvin\ adds substructure to the data that downloads only the explicitly requested data (i.e., a single map or spectrum), minimizing bandwidth and local disk use, if desired.

    \item Individuals must build their own tools to manage the download of data to the local server, which can be complicated to manage efficiently without substantial effort.  Additionally individuals must define a different set of tools for accessing remote data versus local files.  These logistical issues pose a significant barrier to new users.  In contrast, \marvin\ comes with such tools that automatically avoid multiple downloads of the same data.  The same set of \marvin\ tools can be used in different hardware locations (i.e.,  with either local or remote access to the data) with only a single configuration change.

    \item To compare map quantities and spectra, individuals must build their own tools to link spatial locations in the maps to spectra.  In contrast, most of the detailed data access tools are built into \marvin, and internally perform all necessary linkage in a standardized fashion.

    \item Visualizing the data is cumbersome and time-consuming as it requires all data be local, and relies on manual plotting scripts or the repetitive use of third-party tools.  In contrast, \marvin's web interface and Python package provide visualization tools for fast iteration and exploratory analyses.

    \item Individuals' analysis code remains siloed and is not reused.  In contrast, \marvin\ includes some analysis code and serves as a foundation and repository for shared analysis code, which minimizes code duplication across researchers and projects.

\end{itemize}

\marvin\ is structured as a complete ecosystem such that the entire workflow can be performed in a single Python environment or program, but its modular design allows many aspects of \marvin\ to be used independently of each other.  Data can be accessed either through \marvin's provided Tools, or downloaded using \marvin\ but imported and analyzed with other tools.  This flexibility makes \marvin\ a useful tool to a broad range of astronomers.  See Appendix~\ref{app:metals} for an example \marvin\ workflow that examines the metallicities of star-forming spaxels in MaNGA galaxies.

\section{Sustainability}
\label{sec:sustain}

Sustainability is an important part of any software's longevity and usefulness within the community.  Good software is well-documented, easy to maintain, and guided by productive interactions between its users and developers.  Software development is often critically misunderstood to occur in an isolated environment and in static snapshots with little input from the community of users.  Successful software adoption and development depends on easy installation, well-understood code, an open and eager community,  quick identification of breaking changes through continuous testing, and rapid patch updates through continuous deployment.  Below we highlight steps that we have taken to make \marvin\ more sustainable in the long term.

\subsection{Deployment}
\label{sec:deploy}

To streamline installation and reach as wide an audience as possible, we tag and deploy versions of Marvin using the Python Package Index (PyPI)\footnote{\url{https://pypi.org/}}.  Packages on PyPI are installed with the \texttt{pip} package, which simplifies installation by handling the package dependencies.  A \texttt{pip}-installable package simplifies user installation handling all software dependencies automatically.  Installing software packages with a large number of dependencies can interfere with a users' local environment.  To resolve this issue, we also provide full \texttt{conda} environment installations\footnote{\url{https://anaconda.org/sdss/sdss-marvin}}.  For users with the Anaconda distribution of Python, these environments will install \marvin\ in a self-contained virtual environment that does not affect a users' default environment.

\subsection{Open Source Code}
\label{sec:opensource}

SDSS supports and encourages open source software development for all its projects and advocates for a fully transparent software cycle from development to release.  We have adopted the BSD 3-Clause open source software license as it allows for the most complete freedom of use.   The \marvin\ code and its SDSS dependencies are versioned using \texttt{git} and hosted in public Github repositories.  To promote consistency among the SDSS software projects, SDSS encourages all members to adhere to a common set of coding standards and practices\footnote{\url{https://sdss-python-template.readthedocs.io/en/stable/standards.html}}, which we have adopted in the development of \marvin. We additionally tag all versions of our software with a trackable DOI and host it on Zenodo\footnote{\url{https://doi.org/10.5281/zenodo.596700}}.

\subsection{Documentation}
\label{sec:documentation}

Software is only helpful if users understand how to use it, which means providing thorough and well-organized documentation.  \marvin's documentation is structured hierarchically so that it completely covers all of the public-facing code and yet remains easily searchable.  The documentation also contains worked examples that highlight commonly used workflows to associate scientifically-related functionality within the code base.

\marvin's documentation starts with the docstrings in the code, which are included in all of the public-facing code.  Users can access the docstrings on Github to understand the documentation in the context of the code or just simply and quickly view the docstrings while working in an interactive python terminal.  The docstrings also serve as the input for the API reference created by the \texttt{Sphinx} package, which is the basis for web-based documentation pages.

\marvin's web-based documentation\footnote{\url{https://sdss-marvin.readthedocs.io}} is hosted on Read the Docs, which automatically generates the documentation web pages (using \texttt{Sphinx}) every time the code in the ``master'' branch changes on Github.  Using \texttt{Sphinx} with Read the Docs ensures that the documentation is consistent between the code base and the online documentation.  This workflow also enables versioned documentation that corresponds to the versions (tags) of the code, including the data model.  Versioned documentation is critical for understanding how the code base has changed over time and its potential effects on out-dated user code.

An important aspect of the usability of documentation is how quickly users can find answers to their questions.  Read the Docs features a search bar that can be used to find keywords within all of the online documentation.  We also have written documentation pages to help guide users through the code and provide paths for digging deeper into the documentation in useful directions.  The documentation contains use-oriented pages for the major classes and other key aspects, such as installation.  At a higher level, there are various tutorials on topics like first steps, querying, and plotting.  For more complete science examples, there are Jupyter notebooks that are centered around addressing actual science questions.  Finally, there is a cheat sheet that is a one-page quick reference guide to help users remember syntax and function names.

The documentation has benefited tremendously from user feedback within the SDSS community received during in-person workshops, in remote tutorial sessions, and via submitted Github Issues.  Actual user feedback is invaluable, since the developers have a very different perspective that sometimes blinds them to gaps in the documentation.

\subsection{Error Logging}

To better track when errors occur as users utilize \marvin\ in their scientific workflows, all errors are cataloged using Sentry, an error-logging client.  Rather than solely relying on users to report errors, Sentry is an open-source error tracking client that helps developers monitor and fix crashes in real time.  The Sentry client easily integrates into many existing frameworks, including Python and Javascript.  Upon an error within \marvin, Sentry will catalog the error and full traceback along with the relevant software, user, and system information.   Sentry allows us to track when and how often errors occur, automatically create Github Issues, and generally boost the efficiency at which \marvin\ failure modes are identified.

\subsection{Testing}
\label{sec:testing}

Testing is an integral part of developing and maintaining any software package.  Simply put, tests confirm that code executes as intended; however, the benefits extend beyond the confidence inspired by passing tests.  Testing speeds up development by enabling automated checking for regression bugs, where a code change unexpectedly breaks existing functionality, so developers immediately know when a breaking change is introduced.  Testing strongly encourages other good software carpentry practices, such as writing short functions that only do one thing.  Tests significantly lower the risk of refactoring code because they ensure consistent behavior of the code after refactor.  Generally speaking, tests reduce the effort required to maintain a code base.

\subsubsection{Python Testing}

One of our goals is to have complete test coverage of the Python code since it underlies all aspects of \marvin.  \marvin\ uses the \texttt{pytest} package because it requires less boilerplate code relative to the Python standard library package \texttt{unittest}; its multi-level fixtures; the ability to parametrize fixtures; its monkeypatching and mocking options to simulate environments or function calls; and its quick iteration and inspection of test failures.  These features are particularly helpful for ensuring consistent behavior across data releases, analysis and binning types, and data access modes.

\subsubsection{Continuous Integration with Travis-CI}

Software teams are often developing in parallel with one another, and operating on multiple threads at the same time.  This can lead to conflicts and lost time when merging threads back together or when modified or new code introduces breaking changes.  Continuous integration allows a code base to test code changes in a continuous fashion in real time.  \marvin\ uses the Travis-CI system for automated testing.  Travis-CI is free for all projects hosted on Github.  Once a project's repository is connected to Travis, any commit to the code triggers Travis to set up a virtual environment and run the full test suite.  Travis can be configured to run the test suite under different environment conditions, such as different versions of the programming language.

\subsubsection{Code Coverage with Coveralls}

Code coverage is a method for determining how much of the code is covered by the tests.  As the tests run, the lines of code touched by the tests are compared against the total number of lines, and a coverage report is prepared for every code file, and the software package as a whole.  Coveralls is an online interface for visualizing and tracking the code coverage over time, designed to ensure the test coverage of the code is always increasing over time.  It can be connected into existing CI services for real time updates.  After Travis successfully completes a test run, the coverage output is pushed to Coveralls, where developers can see the progress over time, and examine individual files for lines of code untouched by tests.

\subsubsection{Framework Testing}

\texttt{Flask} comes with a test server designed to simulate running a real web server.  It exposes the web and API routes so that tests can be properly written against them.  All back-end web and API routes are fully tested for general compliance and for the desired behavioral outputs given specific users' inputs.   To ensure the web front-end behaves as expected after code changes, we use Selenium combined with the \texttt{Pytest-Flask} extension to simulate a user experience of navigating \marvin\ web pages and interacting with any element on the web page.  These tests can be run locally or hooked into a CI service.  \browserstack\ is a service to run web front-end tests within different versions of multiple browsers, on any operating system, including tablets and mobile devices.  We have connected Travis and \browserstack\ together, so the web front-end tests Travis performs are sent to be run on the \browserstack\ servers.

\section{Summary and Future Potential}
\label{sec:summary}

We have presented the first public release of the \marvin\ software, a suite of tools for interacting with the SDSS-IV MaNGA data set.  We have described the core components of \marvin\ in Section~\ref{sec:coredesign}, the main one being the Multi-Modal Data Access System (Section~\ref{sec:mma}).  The MMA system is a novel method of delivering MaNGA data to the user in an agnostic manner, seamlessly switching between local files and a remote database.  This allows the user to focus more on their scientific analysis with the data and less on the overheads of data access and correct data integration into existing scripts.  The MMA is tightly integrated into a suite of Python Tools (Section~\ref{sec:marvin-tools}) designed to streamline the users' interaction with the core MaNGA data deliverables, and provide remote querying (Section~\ref{sec:query}) against the entire MaNGA data set from within a Python terminal, eliminating the need to download the full sample.  \marvin's \query\ system utilizes a pseudo-natural language syntax parser (Section~\ref{sec:pnls}) to simplify the writing of SQL queries down to the most meaningful component, namely science-pertinent conditionals. \marvin\ provides a web front-end (Section~\ref{sec:marvin-web}), allowing for quick visual exploration of the MaNGA data set, presented in an intuitive, clean manner that lowers the barrier of entry for new users into MaNGA data.  The \marvin\ team of developers is fully committed to an open-source model of development and has adopted modern best practices to ensure long-term software sustainability (Section~\ref{sec:sustain}).  We invite members of the community to adopt the \marvin\ software in their workflows for scientific analysis of MaNGA data, to provide insightful feedback and report issues, or to contribute new features and functionality.

\marvin\ has potential in the future to grow to serve not only the existing community of MaNGA users, but also the overall astronomy community.  In Section~\ref{sec:usecase} we described a typical workflow with the MaNGA data set.  However, this workflow is not really unique to MaNGA.  Replace the MaNGA data with any other astronomical data set and the workflow essentially remains the same.  In the same way that \marvin\ addresses the inherent issues with the MaNGA workflow, a generalization of \marvin\ has the potential to address the same issues for other astronomy workflows.  The most natural extension of \marvin\ is to other surveys producing IFU-based data products, but the design of \marvin\ and the \brain\ inherently make them applicable to generic data sets within astronomy, as well as other scientific disciplines.

\subsection{The \brain}

Astronomy has many logistical challenges and overheads that are often overcome by reinventing existing tools.  There is no suite of tools providing the out-of-the-box capabilities to address these issues.  Many of the underlying Python packages (e.g., \texttt{requests}, \texttt{Flask}, \texttt{SQLAlchemy}) that are often used to solve these challenges are difficult to learn for newcomers and do not provide a usable solution by themselves.  A product is needed that combines these packages in a way that addresses the needs of large scientific data sets with complex interrelationships.  The \brain\ is the beginning of this product.  By further abstracting out \marvin's building blocks into the \brain, the \brain\ will become a complete framework for a data distribution system, providing seamless connections between web components, APIs, and programmatic Python tools.  Furthermore, as the backbone of \marvin, this abstraction will provide a more robust, sustainable framework for \marvin\ that will make it even easier to extend to other data sets.

\subsection{As a Template}

\marvin\ serves as an example of a data distribution system but currently is quite specific to the MaNGA data set with nuggets of generalization scattered throughout it.  With the proper abstraction of the \brain, it will provide a customizable, ``Blank Slate'', template product to handle more complex data sets and become a reusable toolkit for other projects within astronomy and other disciplines.  This product, when given a file and a database presentation of that file, will provide base classes to provide a connected environment surrounding that file, with local and remote access, programmatic tools, a remote query system, and a web front-end to the data with a minimum display view.  Creating a functional application beyond the ``Blank Slate'' would simply involve building a new Python package based off the \brain; subclassing its base classes; and adding functionality and details necessary for their particular application.  This template could be used for rapid deployment and distribution of data sets for any research group ranging in size from small local teams to large collaborations or surveys.

\subsection{\sciserver\ Integration}

While \marvin\ works either as a local analysis package, or for browser-based visual exploration, it still requires local package installation for local analysis, and focuses on single user software usage.  Local package installation can often interfere with custom user environments, while single user usage and analysis limits the ability for collaborative science.  However, an advantage of \marvin\ is that it can be deployed either as a client service, or in a server mode distributing content, into existing archive systems or in different environments.  To enable collaborative and remote scientific analysis, we are in the process of integrating \marvin\ into the \sciserver\ platform.  \sciserver\ \footnote{\url{http://www.sciserver.org/}} is a fully integrated cyber-infrastructure system encompassing related, integrated, tools and services to enable researchers to cope with, and collaborate around, scientific big data.

\sciserver\ integration will enable users to utilize the access and analysis capabilities of \marvin\ without having a local installation.   Collaborative science will be enabled through remote, persistent, \jupyter-notebooks that can be shared amongst multiple users.  Analysis can be offloaded to the \sciserver\ system removing the need for any data to be hosted locally.  Additionally, through the \sciserver\ Compute system, the \marvin\ \query\ tool will be expanded to include asynchronous query capability, allowing intensive queries to be submitted as jobs, similar to the existing SDSS CasJobs system, freeing up the user's local terminal.

\subsection{Community Driven Development}

Our goal with \marvin\ was to make it as easy as possible for users to interact with MaNGA data.  In the pursuit of that goal, we have developed a software framework that is quite extensible, making it extremely easy to contribute tools, features, and functions that enhance \marvin's capabilities.  \marvin\ takes care of all logistical overhead of interacting with the data, as well as defining all core tools that provide most of the desired information.  We realize, however, the greatest usefulness of \marvin\ are the things not yet done.
While the developers have provided the initial scope of \marvin\, the community can greatly expand its usefulness through code contributions.  Through the open source nature of \marvin\ and the ability to wrap the existing suite of tools, the community can easily provide new functionality to aid in the overall scientific usefulness of the MaNGA data set for people at any stage of their scientific career.

\subsection{User Metrics and Broader Impact}
\label{sec:user}

To properly understand the benefit of software within the community, and to better aid in future software funding, it is of paramount importance that software developers play a more active role in assessing the impact and usefulness of their software within the community.  As software itself plays an often overlooked but important role in the pursuit of scientific discovery, developers not only have a role to provide to the community but also to better understand their users for future developments that aid in that discovery.   We aim to address exactly these issues.  Marvin has been operating in Beta for over a year now.  In this time we have been collecting anonymous user statistics and metrics (e.g. web/API route access or query submissions), and obtaining feedback on use cases for users within the collaboration.  With this data, plus the data we will collect with the public release of \marvin, we have a good opportunity to assess the impact of \marvin\ on the community of users.  We hope to provide feedback on the current usage statistics for \marvin, some general examples of scientific workflows utilizing \marvin, as well as lessons learned throughout the development of \marvin.

\acknowledgments

We would like to acknowledge the MaNGA team for supporting this project, helping shape its design, and providing critical feedback and testing during various shakedown phases.  We would like to thank Demitri Muna for his help with \flask\ and database services.  We give thanks to 8485-1901 for its noble sacrifice as our guinea pig.

Funding for the Sloan Digital Sky Survey IV has been provided by the Alfred P. Sloan Foundation, the U.S. Department of Energy Office of Science, and the Participating Institutions. SDSS-IV acknowledges
support and resources from the Center for High-Performance Computing at
the University of Utah. The SDSS web site is www.sdss.org.

SDSS-IV is managed by the Astrophysical Research Consortium for the
Participating Institutions of the SDSS Collaboration including the
Brazilian Participation Group, the Carnegie Institution for Science,
Carnegie Mellon University, the Chilean Participation Group, the French Participation Group, Harvard-Smithsonian Center for Astrophysics,
Instituto de Astrof\'isica de Canarias, The Johns Hopkins University,
Kavli Institute for the Physics and Mathematics of the Universe (IPMU) /
University of Tokyo, the Korean Participation Group, Lawrence Berkeley National Laboratory,
Leibniz Institut f\"ur Astrophysik Potsdam (AIP),
Max-Planck-Institut f\"ur Astronomie (MPIA Heidelberg),
Max-Planck-Institut f\"ur Astrophysik (MPA Garching),
Max-Planck-Institut f\"ur Extraterrestrische Physik (MPE),
National Astronomical Observatories of China, New Mexico State University,
New York University, University of Notre Dame,
Observat\'ario Nacional / MCTI, The Ohio State University,
Pennsylvania State University, Shanghai Astronomical Observatory,
United Kingdom Participation Group,
Universidad Nacional Aut\'onoma de M\'exico, University of Arizona,
University of Colorado Boulder, University of Oxford, University of Portsmouth,
University of Utah, University of Virginia, University of Washington, University of Wisconsin,
Vanderbilt University, and Yale University.

\software{
Anaconda (\url{https://anaconda.org/anaconda/python}),
Astropy \citep[][\url{http://www.astropy.org}]{astropy2013, astropy2018},
Bootstrap (\url{https://getbootstrap.com}),
Browserstack (\url{https://www.browserstack.com})
brain (\url{https://github.com/sdss/marvin_brain}),
Coveralls (\url{https://coveralls.io/}),
D3 (\url{https://d3js.org}),
DyGraphs (\url{http://dygraphs.com}),
FITS \citep{pence2010},
Flask (\url{http://flask.pocoo.org}),
Flask-Login (\url{https://flask-login.readthedocs.io}),
Flask-JWT-Extended (\url{https://flask-jwt-extended.readthedocs.io}),
fuzzywuzzy (\url{https://github.com/seatgeek/fuzzywuzzy}),
git (\url{https://git-scm.com}),
Highcharts (\url{https://www.highcharts.com}),
Jinja2 (\url{http://jinja.pocoo.org/docs}),
JQuery (\url{https://jquery.com}),
Jupyter \citep[][\url{http://jupyter.org}]{kluyver2016},
Matplotlib \citep[][\url{https://doi.org/10.5281/zenodo.61948}]{hunter2007},
networkx (\url{https://networkx.github.io}),
Nginx (\url{https://www.nginx.com}),
OpenLayers (\url{https://openlayers.org}),
pip (\url{https://pypi.org/project/pip}),
Postgres (\url{https://www.postgresql.org}),
pytest (\url{https://docs.pytest.org/}),
Read the Docs (\url{https://readthedocs.org/}),
requests (\url{http://docs.python-requests.org}),
rsync (\url{https://rsync.samba.org}),
sdss-access (\url{https://doi.org/10.5281/zenodo.1410704}),
sdss-tree (\url{https://doi.org/10.5281/zenodo.1410706}),
Selenium (\url{https://www.seleniumhq.org}),
Sphinx (\url{http://www.sphinx-doc.org}),
SQLAlchemy (\url{https://www.sqlalchemy.org}),
sqlalchemy-boolean-search (\url{https://github.com/sdss/sqlalchemy-boolean-search}),
Travis-CI (\url{https://travis-ci.org/}),
uwsgi (\url{https://uwsgi-docs.readthedocs.io})
}

\bibliographystyle{aa}
\bibliography{references}

\appendix

\section{Example Code}

\subsection{Illustration of the \marvin\ MMA}
\label{app:mma}
This illustrates the access of a \marvin\ \cube\ object, utilizing the \marvin\ MMA, in the Python programming language.   After the initial import of the \marvin\ \cube\ tool, a DRP cube is accessible by explicitly providing as input either the full path to a local file or by an object ID.  When explicitly specifying a filename (first example), \marvin\ assumes a \textbf{mode=local} and a \textbf{data\_origin=file}.  When specifying an object ID, \marvin\ will first attempt to open the object locally from a local database or from a file if one is found (second example). If an object cannot be found locally (third example), \marvin\ switches to \textbf{mode=remote} and \textbf{data\_origin=api} to retrieve the object from a remote location.  In each of the cases, the output is the same, an instance of the \marvin\ Cube which wraps the specified DRP datacube.

\begin{lstlisting}[language=Python]
from marvin.tools.cube import Cube
# INFO: No release version set. Setting default to MPL-6

# access locally by filename
c = Cube("/Users/Brian/manga/redux/v2_3_1/8485/stack/manga-8485-1901-LOGCUBE.fits.gz")
# <Marvin Cube (plateifu="8485-1901", mode="local", data_origin="file")>

# access locally by id, will either use a database
c = Cube("8485-1901")
# <Marvin Cube (plateifu="8485-1901", mode="local", data_origin="db")>
# ...or a file
# <Marvin Cube (plateifu="8485-1901", mode="local", data_origin="file")>

# access remotely when no local version found
c = Cube("8485-1902")
# WARNING: local mode failed. Trying remote now.
# <Marvin Cube (plateifu="8485-1902", mode="remote", data_origin="api")>
\end{lstlisting}

\subsection{Map Plotting}
\label{app:map-plotting}

The following code was used to produce the right panel of Figure \ref{fig:datacube}, which a 2$\times$2 multi-panel plot of \Ha\ flux, log(\ntwo\ $\lambda$6585 / \Ha) flux, stellar velocity, and stellar velocity dispersion corrected for instrumental broadening.  This snippet illustrates remote data access for retrieving the maps, map arithmetic and applying logarithms, and the simple but customizable plotting method for \texttt{Map}s.

\begin{lstlisting}[language=Python]
import matplotlib.pyplot as plt
import numpy as np

from marvin.tools import Maps

maps = Maps("7977-12705")

halpha = maps.emline_gflux_ha_6564
nii_ha = np.log10(maps.emline_gflux_nii_6585 / halpha)
stvel = maps.stellar_vel
stsig = maps.stellar_sigma
stsig_corr = stsig.inst_sigma_correction()

fig, axes = plt.subplots(nrows=2, ncols=2, figsize=(12, 11))
halpha.plot(fig=fig, ax=axes[0, 0])
nii_ha.plot(fig=fig, ax=axes[0, 1], title="log([NII]6585 / H-alpha)", snr_min=None)
stvel.plot(fig=fig, ax=axes[1, 0])
stsig_corr.plot(fig=fig, ax=axes[1, 1])
\end{lstlisting}

\subsection{Metallicity Science Use Case}
\label{app:metals}

\begin{figure*}
\centering
\includegraphics[width=\textwidth]{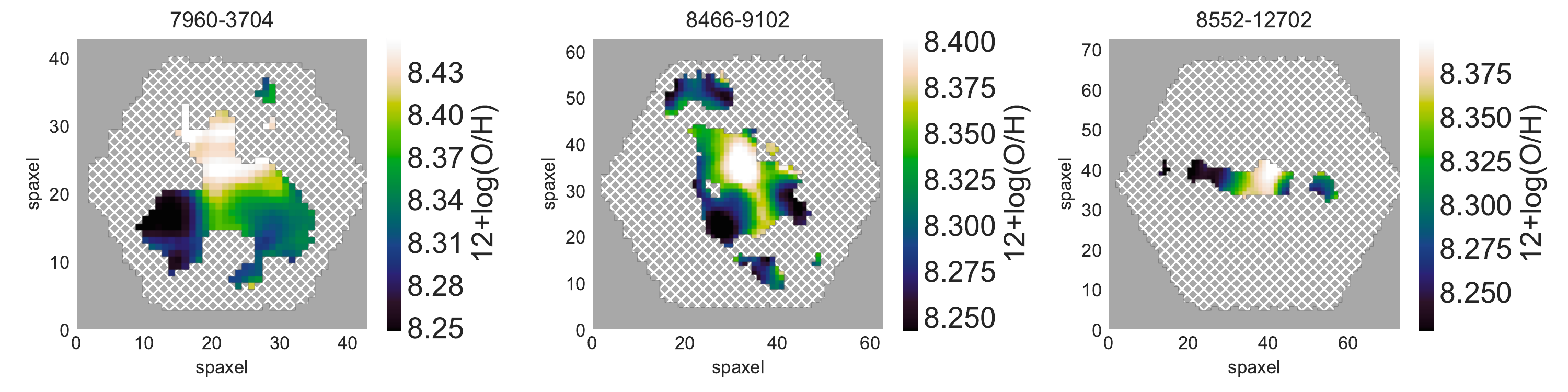}
\caption{Gas-phase metallicity maps for three galaxies (7960-3704, 8466-9102, and 8552-12702) with only star-forming spaxels shown.}
\label{fig:metallicity_maps}
\end{figure*}

\begin{figure*}
\centering
\includegraphics[width=\textwidth]{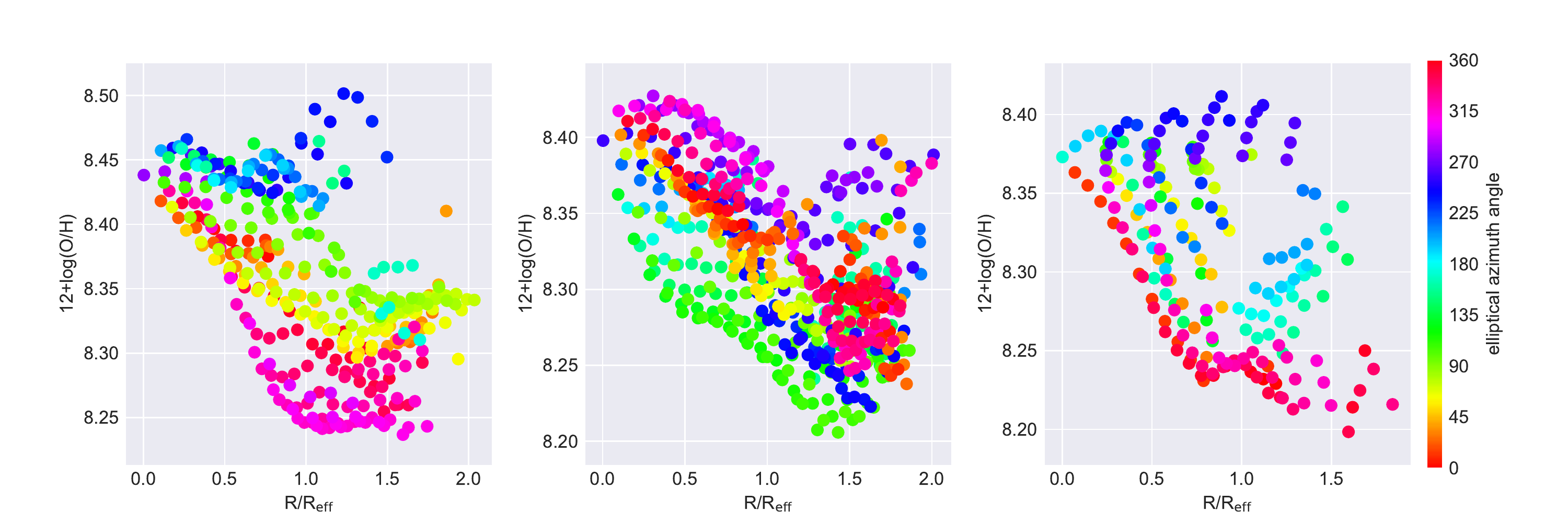}
\caption{Radial gas-phase metallicity gradients (in units of effective radius $R_\mathrm{eff}$) for three galaxies (7960-3704, 8466-9102, and 8552-12702) color-coded by elliptical azimuthal angle with only star-forming spaxels shown.}
\label{fig:metallicity_gradients}
\end{figure*}

The following code computes and plots gas-phase metallicity maps (Figure \ref{fig:metallicity_maps}) and gradients (Figure \ref{fig:metallicity_gradients}) for star-forming spaxels only.  The code selects galaxies based on global parameters using a \marvin\ query.  It then generates masks to select the star-forming spaxels using \marvin's BPT tool (see Section \ref{sec:utilities}).  Finally it takes advantage of \marvin's map arithmetic to compute metallicities.

\begin{lstlisting}[language=Python]
from matplotlib import cm
import matplotlib.pyplot as plt
import numpy as np

from marvin.tools.query import Query
import marvin.utils.plot.map as mapplot

# Define query and run it
searchfilter = ("nsa.sersic_logmass >= 9 and "
                "nsa.sersic_logmass <= 9.1 and "
                "nsa.elpetro_mag_g_r < 0.3")
query = Query(searchfilter=searchfilter, limit=3)
results = query.run()

# Convert query results into a list of Marvin data product objects
results.convertToTool("maps")
galaxies = results.objects

metallicity = []
for it in galaxies:

    # Lazy-load [NII] and Halpha maps for each galaxy
    nii = it.emline_gflux_nii_6585
    halpha = it.emline_gflux_ha_6564

    # Compute metallicity, 12+log(O/H), using Eq. 1 from Pettini & Pagel (2004)
    N2 = np.log10(nii / halpha)
    oh = 8.90 + 0.57 * N2
    metallicity.append(oh)

# Create BPT masks for each galaxy
bpt_masks = [it.get_bpt(return_figure=False, show_plot=False) for it in galaxies]

# Default bad data labels for MANGA_DAPPIXMASK
labels_bad = ["NOCOV", "UNRELIABLE", "DONOTUSE"]

# Apply BPT cut to select star-forming spaxels and mask out spaxels with bad data
masks = []
for metal, bpt in zip(metallicity, bpt_masks):
    # Mask out non-star-forming spaxels
    starforming = bpt["sf"]["global"]
    non_sf = ~starforming * metal.pixmask.labels_to_value("DONOTUSE")

    # Merge the non-star-forming mask with the MANGA_DAPPIXMASK mask
    masks.append(metal.mask | non_sf)

# Metallicity maps
fig, axes = plt.subplots(1, 3, figsize=(12, 3))
for ax, metal, mask, gal in zip(axes, metallicity, masks, galaxies):
    mapplot.plot(dapmap=metal, mask=mask, fig=fig, ax=ax,
                 cblabel="12+log(O/H)", title=gal.plateifu)

fig.tight_layout()

# Metallicity gradients
fig, axes = plt.subplots(1, 3, figsize=(12, 4))
plt.subplots_adjust(left=0.08, wspace=0.25)
for ax, metal, mask, gal in zip(axes, metallicity, masks, galaxies):
    mappable = ax.scatter(
        gal.spx_ellcoo_r_re.value[~mask.astype(bool)],
        metal.value[~mask.astype(bool)],
        c=gal.spx_ellcoo_elliptical_azimuth.value[~mask.astype(bool)],
        cmap=cm.hsv,
        vmin=0,
        vmax=360,
    )
    ax.set_xlabel("R/R$_\mathrm{eff}$")
    ax.set_ylabel("12+log(O/H)")
    ax.set_title(gal.plateifu)

cax = fig.add_axes([0.91, 0.105, 0.01, 0.78])
ticks = np.linspace(0, 360, 9)
cb = fig.colorbar(mappable, cax, ticks=ticks)
cb.set_label("elliptical azimuth angle")
\end{lstlisting}

\section{\marvin\ tutorials}

In addition to installation instructions and detailed descriptions of the \marvin\ components, the online documentation also contains tutorials\footnote{\url{https://sdss-marvin.readthedocs.io/en/stable/tutorials.html}}, including exercises on specific science cases and example \jupyter\ notebooks.  The tutorials are intended to familiarize the user with working with MaNGA data via \marvin.  Each tutorial is designed as a series of steps that build on top of one another and flow together to provide a layered learning experience, guiding the user from a straightforward approach to more in-depth \marvin\ usage.  These tutorials act as a good entry point for new users diving into MaNGA data.

A few example \marvin\ tutorials are:

\begin{itemize}
\item Lean Tutorial\footnote{\url{https://sdss-marvin.readthedocs.io/en/stable/tutorials/lean-tutorial.html}}: an example project using \marvin\ from start to finish.  It focuses on the calculation of the \ntwo/\Ha\ ratio for star-forming spaxels in galaxies with stellar mass log($M_\star$) ~=~10--11 M$_\sun$.

\item Plotting Tutorial\footnote{\url{https://sdss-marvin.readthedocs.io/en/stable/tutorials/plotting-tutorial.html}}: plotting examples of increasing complexity from a single map or spectrum to heavily customized multi-panel map plots.

\item Sample Selection Tutorial\footnote{\url{https://sdss-marvin.readthedocs.io/en/stable/tutorials/sample-selection.html}}: examples of how to search on the Primary, Secondary, and Color-Enhanced samples that make up the MaNGA main sample \citep{yan2016b, wake2017}.
\end{itemize}

%
%
%
%
%

Additional \marvin\ tutorials are available on the main SDSS DR15 site\footnote{\url{https://www.sdss.org/dr15/manga/manga-tutorials/marvin-tutorial/}}.  These tutorials are designed as simple introductions to a few basic features of \marvin, with more advanced tutorials in the main \marvin\ documentation.  Tutorials are provided for both \marvin's front-end web interface, as well as the Tools.  The Tools tutorials focus on  accessing spectra from individual spaxels, plotting maps for derived analysis properties with custom masks, identifying unique bins, and extracting a binned spectrum.  For the web tutorials, there are two videos provided to highlight some relevant features.  One video focuses on exploring an individual MaNGA galaxy.  It highlights basic information, the dynamic interactive maps and spectral visualization, and detailed information from the NSA catalog.  The second video focuses on performing queries on the MaNGA dataset.  It explains how to use the simplified SQL interface, the guided SQL builder, and the table of search results.  Both video-tutorials are subtitled in English.

\end{document}